\documentclass[review]{elsarticle}[12pt]

\usepackage{lineno,hyperref}
\usepackage{graphicx}
\usepackage{caption}
\usepackage{subcaption}

\makeatletter
\def\ps@pprintTitle{%
  \let\@oddhead\@empty
  \let\@evenhead\@empty
  \def\@oddfoot{\reset@font\hfil\thepage\hfil}
  \let\@evenfoot\@oddfoot
}
\makeatother

\begin{document}

\begin{frontmatter}

\title{On nonlinearity implications and wind forcing in Hasselmann equation}

\author[a2,a3,a4]{Pushkarev Andrei \fnref{fn1} }

\author[a1,a2,a3,a4]{Zakharov Vladimir}

\address[a1]{University of Arizona, 617 N. Santa Rita Ave., Tucson, AZ 85721, USA}
\address[a2]{LPI, Leninsky Pr. 53, Moscow, 119991 Russia}
\address[a3]{Pirogova 2, Novosibirsk State University, Novosibirsk, 630090 Russia}
\address[a4]{Waves and Solitons LLC, 1719 W. Marlette Ave., Phoenix, AZ 85015 USA}

\fntext[fn1]{Corresponding author, \textit{E-mail}: dr.push@gmail.com}

\begin{abstract}
We discuss several experimental and theoretical techniques historically used for Hasselmann equation wind input terms derivation. We show that recently developed $ZRP$ technique in conjunction with high-frequency damping without spectral peak dissipation allows to reproduce more than a dozen of fetch-limited field experiments. Numerical simulation of the same Cauchy problem for different wind input terms has been performed to discuss nonlinearity implications as well as correspondence to theoretical predictions. 

\end{abstract}

\begin{keyword}
\texttt{Hasselmann equation, wind-wave interaction, wave-breaking dissipation, nonlinear interaction, self-similar solutions, Kolmogorov-Zakharov spectra}
\end{keyword}

\end{frontmatter}


\section{Introduction}

\noindent It is generally accepted nowadays that ocean surface wave turbulence is described by Hasselmann equation (hereafter $HE$)  

\begin{equation}
\label{HE}
\frac{\partial \varepsilon }{\partial t} +\frac{\partial \omega _{k} }{\partial \vec{k}} \frac{\partial \varepsilon }{\partial \vec{r}} =S_{nl} +S_{in} +S_{diss}
\end{equation}

\noindent for spectral energy density $\varepsilon =\varepsilon (\vec{k},\vec{r},t)$,  wave dispersion $\omega =\omega (k)$ and nonlinear, wind input and wave-breaking dissipation terms $S_{nl}$, $S{}_{in}$ and  $S_{diss}$  correspondingly.

\noindent While this acceptance implicitly assumes that $HE$ is some sort of mathematical reduction of primordial Euler equations for incompressible fluid with free surface, it is formally true, in fact, only for advection $\frac{\partial \omega _{k} }{\partial \vec{k}} \frac{\partial \varepsilon }{\partial \vec{r}} $ and four-wave interaction $S_{nl} $ terms.

\noindent As far as concerns $S_{in}$ and $S_{diss}$ terms, there is no consensus in the worldwide oceanographic community about their parameterization. To our belief, it is one of the reasons, indeed, for ``tuning knobs'' (adjusting coefficients) necessity in operational models for their adjustment to different ocean situations set-ups.

\noindent Another reason for using ``tuning knobs'' is the underestimation of the leading role of $S_{nl}$. In other words, the role of the ``tuning knobs'' also consists in ``undoing'' the deformation incurred to the primordial equations model through substitution of the exact nonlinear term $S_{nl}$ with $DIA$-like simplifications.

\noindent We believe that such currently widely accepted approach of using "tuning knobs" is conceptually misleading, and continuing efforts on improving the source terms are fruitless because the nonlinear term $S_{nl}$ is the leading term of $HE$ \cite{R1}, \cite{R2}. All other source terms are, in a sense, relatively small corrections in significant ranges of frequency.
 
\noindent Dominations of the nonlinearity exhibit itself in $HE$ in the form of self-similar solutions, observed in the field and numerical experiments \cite{R3,R4,R5,R6,R7,R8}.

\noindent Self-similarity analysis in conjunction with the field and numerical experiments analysis allowed to build new $ZRP$ wind input term \cite{R9} -- analytical solution of $HE$.

\noindent In the current paper we use alternative approach to $HE$ simulation, which in addition to new $ZRP$ wind input term uses another physically based assumptions -- absence of the spectral peak dissipation and implicit wave damping due to wave-breaking. We justify this new approach through comparison with more than the dozen of field measurements.

\noindent Another result of current paper is the development of the set of tests, based on self-similarity analysis, allowing to make the judgment relative to correctness of arbitrary numerical simulation performed in the frame of $HE$ through comparison of the observed characteristics of wave ensemble with theoretical prediction.

\section{Current state of wind input source terms}

\noindent Nowadays, the number of existing models of $S_{in}$ is large, but neither of them have firm theoretical justification. Different theoretical approaches argue with each other. Detailed description of this discussion can be found in the monographs \cite{R10}, \cite{R11} and the papers  \cite{R12}, \cite{R13}, \cite{R14}, \cite{R15}, \cite{R16}.

\noindent The development of wind excited waves models has begun as far back as 20-ies of the last century in the well-known works of Jeffreys \cite{R17}, \cite{R18}. His model is semi-empiric and includes unknown "sheltering coefficient". All other existing theoretical models are also semi-empiric, with one exclusion -- famous Miles model \cite{R16}. This model is rigorous, but is related to idealized situation -- initial stage of waves excitation by laminar wind with specific wind profile $U(z)$.

\noindent Miles theory application is hampered by two circumstances. First is the fact that atmospheric boundary layer is the turbulent one, and creation of rigorous analytical theory of such turbulence is nowadays unsolvable problem.

\noindent There is the opinion, however, that wind speed turbulent pulsations are small with respect to horizontal velocity $U(z)$, and they can be neglected. This doesn't mean that turbulence is not taken into account at all. It is suggested that the role of the turbulence consists in formation of the averaged horizontal velocity profile. 

\noindent This widely spread opinion is that horizontal velocity profile is distributed by logarithmic law
\begin{equation}
\label{LogProfile}
U(z) = 2.5 u_* \ln{}\frac{z}{z_0}
\end{equation} 
\noindent Here $u_*$ is friction velocity and $z_0$ -- the roughness parameter
\begin{equation}
\label{Charnok}
z_0 = C_{ch} \frac{u_*^2}{g}
\end{equation} 
\noindent where $C_{ch} \simeq 10^{-2}$ is experimental dimensionless Charnok constant.

\noindent One should note that appearance of anomalously small constants, not having "formal justification", is extremely rare phenomenon in physics. Eq. (\ref{LogProfile}), (\ref{Charnok}) mean that roughness parameter is very small: for typical ocean conditions -- wind speed $10$ m/sec on the height $z=10 \,\, m$ we get $z_0 \simeq 10^{-4} \,\, m$. Such roughness is only twice the size of viscid layer, defined from multiple experiments on turbulent wind flow over smooth metal plates.

\noindent Usage of Eqs.(\ref{LogProfile}), (\ref{Charnok}) assumes therefore that ocean behaves as smooth metal surface. This is not correct. Horizontal momentum is transferred to the smooth plate on it's surface itself, while in the ocean this process happens differently.

\noindent Momentum offtake from atmospheric boundary layer is smoothly distributed over the whole width of the boundary layer and begins from the highest "concurrence layer", i.e. from the height where phase speed of the fastest wave matches the horizontal velocity.

\noindent Momentum offtake leads to horizontal velocity distribution $U(z)$ dependence on time, waves development level and energy spectrum. Meanwhile, Miles instability increment is extremely sensitive to the horizontal velocity profile (there is no waves excitation for linear profile $U(z)$  for Miles theory, for example). The velocity profile is especially important for slight elevations of the order of several centimeters over the water surface, which is almost unknown and difficult for experimental measurements. However, there are some advances in this direction \cite{R19,R20}.

\noindent The necessity of taking into account of waves feedback on the horizontal velocity profile has been understood long time ago in the works of Fabrikant \cite{R21} and Nikolaeva, Zyrling \cite{R22} later continued in the works of Jannsen \cite{R23} and explained in details in the monograph \cite{R11} in the form of "quasi-laminar" theory. This theory is not accomplished.

\noindent To consider the theory as self-consistent even in the approximation of turbulence absence, it is necessary to solve equations describing horizontal velocity profile $U(z)$ together with Hasselmann equation, describing energy spectrum evolution. This is not done yet.

\noindent Aside that fact, many theoreticians do not share share the opinion about turbulent pulsations insignificance, and consider them as the leading factor. Corresponding $TBH$ theory by Townsend, Belcher and Hunt \cite{R12} is alternative to quasi-laminar theory. Both theories are discussed in \cite{R24}.

\noindent There is another approach, not connected with experimental analysis -- numerical simulation of boundary atmospheric layer in the frame of empiric theories of turbulence. It was developed in the works \cite{R13,R14,R15,R25}. Since those theories are insufficiently substantiated, the same relates to derived wind input terms.

\noindent For all the variety of theoretical approaches of $S_{in}$ definitions, all of them are are "quasi-linear", assuming:

\begin{eqnarray*}
S_{in} = \gamma(\omega, \phi) \varepsilon(\omega, \phi)
\end{eqnarray*} 

\noindent where standard relation

\begin{eqnarray*}
\gamma(\omega, \phi) = \frac{\rho_a}{\rho_w} \omega \beta(\frac{\omega}{\omega_p}, \phi)
\end{eqnarray*} 

\noindent is being used. Here $\omega_p = \frac{g}{u}$, where $u$ is the wind speed, defined differently in individual models. Function $\beta$ is dimensionless and is growing with the growth of $\frac{\omega}{\omega_p}$.

\noindent However, even for  the most "aggressive" models of wind input terms the value of $\beta$ does not exceed several units, but usually $0 < \beta < 1$. In some models (see,for example \cite{R15}) $\beta$ becomes negative for the waves propagating faster than the wind, or under large angle with respect to the wind.

\noindent Looking at multiple attempts of $S_{in}$ experimental definition, one should note that all of them should be carefully critically analyzed. That criticism is not about the integrity of measurements itself, but about the used methodology and data interpretation correctness, and the possibility of transfer of the conclusions made in artificially created environment to real ocean conditions.

\noindent Another significant amount of experiments, belonging to so-called "fractional growth method" category, has been performed through energy spectrum measurement in time and calculation of the corresponding $\gamma$ through

\begin{eqnarray}
\label{FracGrowth}
\gamma(\omega, \phi) = \frac{1}{\varepsilon(\omega,\phi)} \frac{\partial \varepsilon(\omega, \phi)}{\partial t}
\end{eqnarray} 

\noindent Eq.(\ref{FracGrowth}) is, in fact, the linear part, or just two terms of the $HE$ Eq.(\ref{HE}). This method is intrinsically wrong, since it assumes that either advection $\frac{\partial \omega _{k} }{\partial \vec{k}}\frac{\partial \varepsilon }{\partial \vec{r}}$ and nonlinear $S_{nl}$ terms of Eq.(\ref{HE}) are absent together, or relation 

\begin{eqnarray}
\label{LimFetch}
\frac{\partial \omega }{\partial \vec{k}} \frac{\partial \varepsilon }{\partial \vec{r}} =S_{nl}
\end{eqnarray} 

\noindent is fulfilled. 

\noindent First assumption is simply not correct, since neglected terms are defining in ocean conditions. The second assumption is also wrong since Eq.(\ref{LimFetch}) contradicts Eq.(\ref{FracGrowth}). Therefore, in the relation to "fractional growth method" we are just citing the single relevant publication by Plant \cite{R26} where, it seems, author well understood the scarcity of the "fractional growth method".

\noindent As a matter of fact, the real interest present the experiments, which used measurements of the correlation between the speed of the surface growth and the pressure to the surface:

\begin{eqnarray}
Q(\omega) = Re <\eta(\omega) P^*(\omega)> 
\end{eqnarray}

\noindent Unfortunately, the number of such experiments is limited, and not all of them have significant value for ocean phenomena description. Also, one should take out of consideration the experiments performed in laboratory conditions.

\noindent Consider, for example, the set of experiments described in \cite{R27}. These experimenst were performed in the wave tank of $40 \,\, m$  length and $1 \,\, m$ depth. Experimentators created the wind blowing at the speed up to $16 \,\, m/sec$, but they studied only short waves no longer than $3 \,\, m$, moving no faster than $1.3 \,\, m/sec$. Therefore, they studied the very short-wave tail of the function $\beta$ in the conditions far from the ocean ones. The value of these measurements is not significant.

\noindent The same arguments relate to multiple and precisely performed measurements in the Lake George, Australia \cite{R28}. The depth of this lake, in average, is about $1\,\, m$. That is why on its surface can propagate the waves not faster than $3.3 \,\, m/sec$. The typical wind speed, corresponding these measurements, was $8-12 \,\, m/sec$. 

\noindent Therefore, while the results of these measurements are quite interesting and correspond to theoretical predictions \cite{R29}, obtained expression for $S_{in}$ is quite arguable, not only because is non-improvement to "quasi-linear" theory, but also being in complete contradiction with it.

\noindent After critical analysis of experiments on $S_{in}$ measurements, only three of those deserve an attention. Those are the experiments by Snyder et al. \cite{R30}, Hsiao, Shemdin \cite{R31} and Hasselmann, Bosenberg \cite{R32}. The experiments were performed in the open ocean and measured direct correlations of surface speed change and the pressure.

\noindent Those experiments were performed fairly long time ago, their accuracy is not quite high and scatter of data is significant. Therefore, their interpretation is quite ambiguous. Anyhow, these experiments produced two well-known formula for $\beta$. For Snyder and Hasselmann, Bosenberg experiments:

\begin{eqnarray}
\beta &=& 0.24 (\xi-1), \,\,\, \xi=\frac{\omega}{\omega_0} \cos{\phi} \\
\beta &=& 0, \,\,\, \xi < 1
\end{eqnarray}

\noindent and for Hsiao-Shemdin

\begin{eqnarray}
\beta = 0.12 (0.85 \xi-1)^2 \\
\beta = 0 \,\,\, otherwise
\end{eqnarray}

\noindent The difference between these $S_{in}$ formula is significant. Comparison of wind forcing performed on measured spectra \cite{R5} shows that Snyder-Hasselmann-Bosenberg form gives $5 \div 6$ times bigger value of $S_in$ than Hsiao-Shemdin one.

\noindent Furthermore, the Hsiao-Shemdin form agrees with Jeffreys theoretical model, while  Snyder-Hasselmann-Bosenberg one is in disagreement with any known theoretical models.

\noindent Summing up, we can conclude that at the moment there is no solid parameterization of $S_{in}$ accepted by worldwide oceanographic community. Keeping that fact in a mind, we decided to go our own way -- not to build new theoretical models and not to reconsider both old and completely new measurements of $S_{in}$.

\noindent For almost seventy year, counting from works of Garrett and Munk \cite{R53}, physical oceanography assimilated tremendous amount of experimental facts on basic wind-wave characteristics -- wave energy and spectral peak frequency as a functions of limited fetch.  Such experiments are described in works \cite{R5,R6,R7,R8}. From the other side, numerical methods of Hasselmann equation \ref{HE} solution for exact term $S_{nl}$ and specified in advance terms $S_{in}$ and $S_{diss}$ have been improved significantly. This can be done not only for duration-limited domain, but for fetch-limited domain too.

\noindent Therefore, we proposed purely new pragmatic approach to definition of $S_{in}$. We have chosen $S_{in}$ function in such a way that numerical solution of Hasselmann equation explains maximum amount of known field experiments.The result was the $S_{in}$ function described in details in \cite{R4} and named thereafter $ZRP$ function.

\noindent It is important to emphasize that work \cite {R4} assumed localization of energy dissipation in short waves. This assumption contradicts widely accepted concept, but we explain the difference in the following chapter.

\section{Two scenarios of wave-breaking dissipation term: spectral peak or high-frequency domination?}

\noindent In current section we explain why there is no need to use dissipation in the spectral peak area.

\noindent The spectral peak frequency damping is widely accepted practice, and is included as an option in the operational models $WAM$, $SWAN$ and $WW3$. Historically, it was done apparently by need. 

\noindent The necessity was caused by wind input function $S_{in}$ in Snyder form. Fast wave energy growth was observed in no-dissipation calculations, which didn't match results of field measurements. Despite the result was obtained with the help of $DIA$ model of $S_{nl}$, it is qualitatively correct, because is also confirmed by our numerical calculations using exact nonlinear interaction term $S_{nl}$.

\noindent It is shown below that Snyder wind input without dissipation gives $5-6$ times bigger energy growth than other tested wind input functions ($ZRP$, Tolman-Chalikov and Hsiao-Shemdin). This doesn't mean, however, that long-wave dissipation exist, indeed. From our viewpoint, the necessity of its introduction is explained by Snyder model imperfection, based on not quite accurate experiments.

\noindent We don't see the physical reasons for energy-containing long waves breakings. Their steepness in the conditions of typically developed wave turbulence is not big: $\mu = <\nabla \eta^2>^{1/2} \sim 0.1$, or even smaller. Because this value is very far from limiting steepness of Stokes wave $\mu_S \simeq 0.3$, thise waves are essentially weakly-nonlinear. Besides those waves, more shorter waves inevitably develop, having the steepness approaching to the critical one, and those waves break. There is nevertheless no reason to expect that these waves have the same phase velocity as the energy-containing ones.

\noindent Unfortunately, the theory of "wave-breakers" is not developed yet. In our view, which we don't consider based enough, one of the possible variants of such theory could be the following.

\noindent The primordial Euler equations for potential flow of deep fluid with free surface has the self-similar solution

\begin{eqnarray}
\eta (x,t)=gt^{2} F\left(\frac{x}{gt^{2} } \right)
\end{eqnarray}
 
\noindent This solution was studied numerically in the framework of simplified $MMT$ ($Maida-McLaughlin-Tabak$) model of Euler equations \cite{R33}.

\noindent In Fourier space this solution describes the propagation to high wave-numbers and returning back to dominant wave spectral peak of fat spectral energy tail, corresponding in real space to sharp wedge formation at time $t=0$ and space point $x=0$. This solution describes formation of the "breaker".

\noindent In the absence of dissipation, this event is invertible in time. Presence of high-frequency dissipation chops off the end of the tail, just like ``cigar cutter", and violates the tail invertability. Low and high harmonics, however, are strongly coupled in this event due to strong nonlinear non-local interaction, and deformed high wave-numbers tail is almost immediately returns to the area of spectral peak.  As soon as fat spectral tail return to the area of the spectral peak, total energy in the spectrum diminishes, which causes settling of the spectral peak at lower level of energy. This process of "shooting" of the spectral tail toward high wave-numbers, and its returning back due to wave breaking is the real reason of "sagging down" of the energy profile in the spectral peak area, but was erroneously associated with the presence of the damping in the area of spectral peak.

\noindent This explanation shows that individual wave-breakings studies \cite{R29}, \cite{R34} are not the proof of spectral peak damping presence.

\noindent Also there is another, direct proof of the fact that the damping is localized in the area of short waves. It is the measurements of quasi-one-dimensional "breakers" speed propagation -- strips of foam, which accompany any developed wave turbulence. Those airplane experiments, recently performed by P.Hwang and his team \cite{R35,R36,R37,R38}, show that wave breakers propagate 4-5 times slower than crests of leading waves. 

\noindent Based on the above discussion, we propose to use only high-frequency damping as a basis of alternative framework of $HE$ simulation. One can ``implicitly'' insert this damping very easily without knowing its analytic form via spectral tail continuation by Phillips law $\sim \omega^{-5}$.

\noindent Replacement of high-frequency spectrum part by Phillips law is not our invention.  It is the standard tool offered as an option in operational wave forecasting models, known as the "parametric tail", and corresponds to high-frequency dissipation, indeed. For the practical definition of Phillips tail it's necessary to know two more parameters: coefficient in front of it and starting frequency. 

\noindent The coefficient in front of $\omega ^{-5} $ is not exactly known, but is unnecessary to be defined in the explicit form -- it is dynamically determined from the continuity condition of the spectrum. As far as concerns another unknown parameter -- the frequency where Phillips spectrum starts -- we define it as $f_0 =\frac{\omega_0}{2\pi } = 1.1$ Hz as per Resio and Long experimental observations \cite{R39,R40}. 

\noindent That is the way the high frequency implicit damping is incorporated into alternative computational framework of $HE$. We think that the question of finer details of high-frequency damping structure is of secondary importance at current stage of alternative framework development.

\section{Checking of the new modeling framework against theoretical predictions and field measurements}

\noindent To check alternative framework for $HE$ simulation, we performed numerical tests for waves excitation in limited fetch conditions. As it was already mentioned, alternative framework is based on exact nonlinear term $S_{nl}$ in $WRT$ form and $ZRP$ new wind input term:

\begin{eqnarray} 
&&S_{in} = \gamma \varepsilon \\
&&\gamma = 0.05 \frac{\rho _{air} }{\rho _{water} } \omega \left(\frac{\omega }{\omega _{0} } \right)^{4/3} f(\theta ) \label{ZRP1} \\
&&f(\theta ) = \left\{\begin{array}{l} {\cos ^{2} \theta {\rm \; \; for\; -}\pi {\rm /2}\le \theta \le \pi {\rm /2}} \\ {0{\rm \; \; otherwise}} \end{array}\right. \\
&&\omega _{0} = \frac{g}{u_{10} }, \,\,\,  \frac{\rho _{air} }{\rho _{water} } =1.3\cdot 10^{-3}
\end{eqnarray} 

\noindent The coefficient $0.05$ in front of Eq.(\ref{ZRP1}) was found through carefully performed numerical experiments with different coefficient values to get the best correspondence with experimental data.

\noindent Thus

\begin{equation}
\beta = 0.05 \omega \left(\frac{\omega }{\omega _{0} } \right)^{4/3} f(\theta )
\end{equation}

\noindent We have chosen $\beta$ in the form of power function of frequency for the following reasons. It is well known from various field experiments \cite{R6} that wave energy and spectral frequency maximum dependencies on the fetch are the power functions of fetch:

\begin{eqnarray}
\label{Pdep}
\varepsilon &= \varepsilon_0 \chi^{p}  \\
\label{Qdep}
\omega &= \omega_0 \chi^{-q} 
\end{eqnarray} 

\noindent This observation is in excellent agreement with the fact that conservative stationary Hasselmann equation

\begin{equation}
\frac{\partial \omega}{\partial k} \frac{\partial \varepsilon}{\partial x} =S_{nl}
\end{equation}

\noindent has two-parameter family of self-similar solutions \cite{R3,R4,R5,R6,R7,R8}:

\begin{equation} 
\varepsilon = \chi^{p+q} F(\omega \chi^{q} ) \label{GrindEQ__3_} 
\end{equation}

\noindent which lead to dependencies Eqs.(\ref{Pdep}), (\ref{Qdep}).

\noindent One of the $p, q$  is free, but they are connected by relation

\begin{equation}
\label{MagicLaw}
10p - 2q = 1
\end{equation}

\noindent called thereafter the "magic law".

\noindent Analysis of field experiments \cite{R6} shows that "magic law" is fulfilled with high accuracy in many of them.

\noindent $HE$ Eq.(\ref{HE}) has self-similar solutions only if it is supplied by wind input term in power form:
\begin{equation}
\beta = \omega^s f(\theta)
\end{equation}
\noindent and self-similar substitution  gives in this case  $q=\frac{1}{2+s}$. In the most known experiments $p=1$ and $q=3/10$. That gives us the value $s=4/3$.

\noindent It is important to note that if $\beta(\frac{\omega}{\omega_p})$ is not the power function, the Eq.(\ref{HE}) solution in the absence of long-wave dissipation is "quazi-self-similar" in typical cases. In this case $p$ and $q$ are slow functions of the fetch, but the "magic law" Eq.(\ref{MagicLaw}) is still fulfilled \cite{R8}. Strictly speaking, an existence of the universal for any ocean conditions expression of $\beta(\omega,\theta)$ is not proved, because turbulent boundary layer is different for typical oceans, passats (trade winds) or mountain coastal line. Therefore, the fact of explanation by $ZRP$ expression for $S_{in}$ of at least half of known field experiments can be considered as big success.

\begin{figure}
        \centering
        \begin{subfigure}[b]{0.45\textwidth}
                \includegraphics[width=\textwidth]{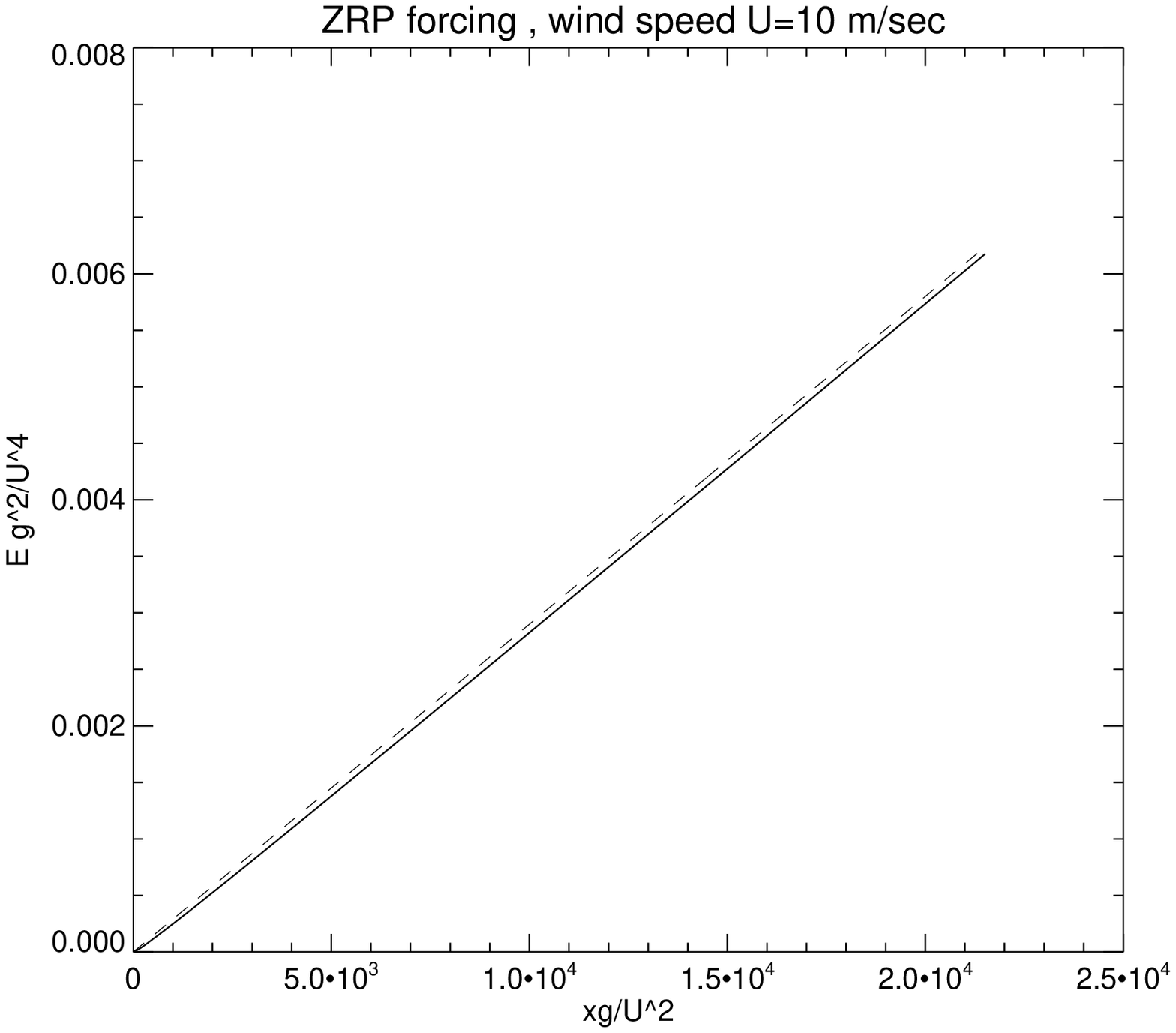}
                \caption{Solid line - numerical experiment, dashed line -- fit by $2.9\cdot 10^{-7} \cdot \frac{xg}{U^{2}}$ }
                \label{AltFormA}
        \end{subfigure}
\qquad
        \begin{subfigure}[b]{0.45\textwidth}
                \includegraphics[width=\textwidth]{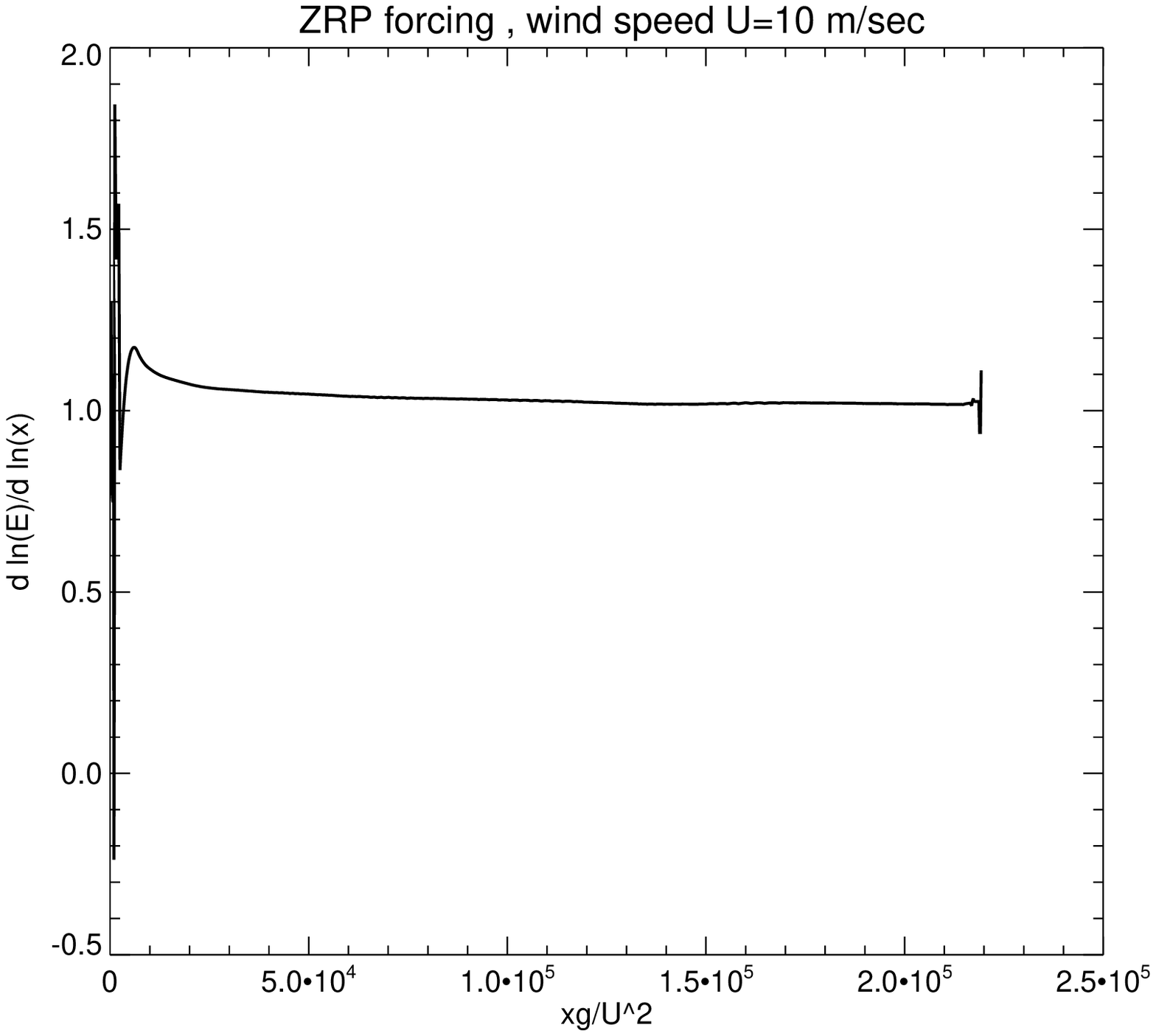}
                \caption{Exponent $p$ of the energy growth as the function of fetch $x$}
                \label{AltFormB}
        \end{subfigure}
      \caption{}\label{AltForm}
\end{figure}

\noindent Fig.\ref{AltForm} shows that total energy is growing along the fetch by power law in accordance with Eq.(\ref{Pdep}) with $p=1.0$.


\begin{figure}
        \centering
        \begin{subfigure}[b]{0.45\textwidth}
                \includegraphics[width=\textwidth]{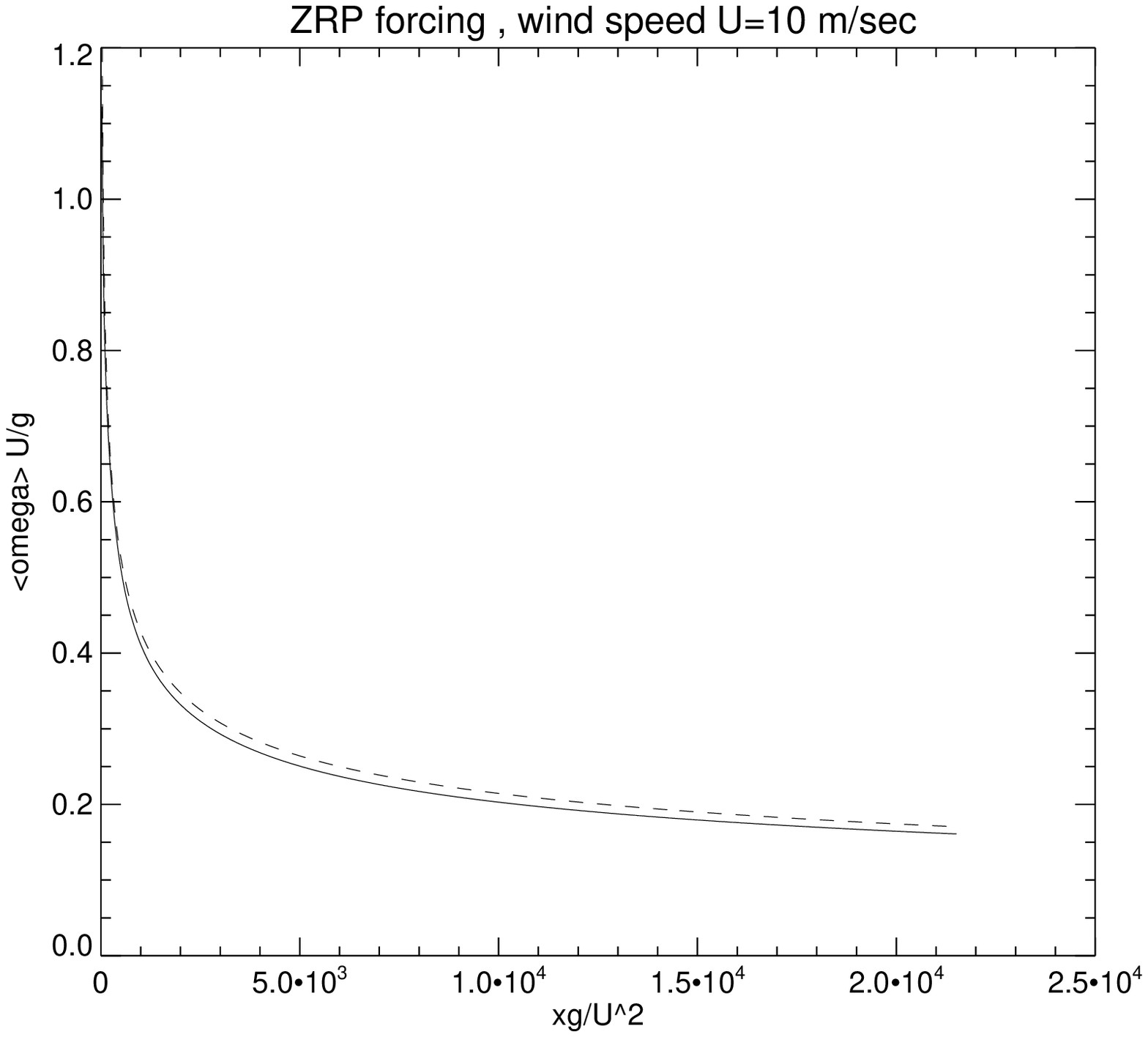}
                \caption{Solid line - numerical experiment, dashed line -- fit by $3.4\cdot \left(\frac{xg}{U^{2} } \right)^{-0.3}$. }
                \label{MeanFreqA}
        \end{subfigure}
\qquad
        \begin{subfigure}[b]{0.45\textwidth}
                \includegraphics[width=\textwidth]{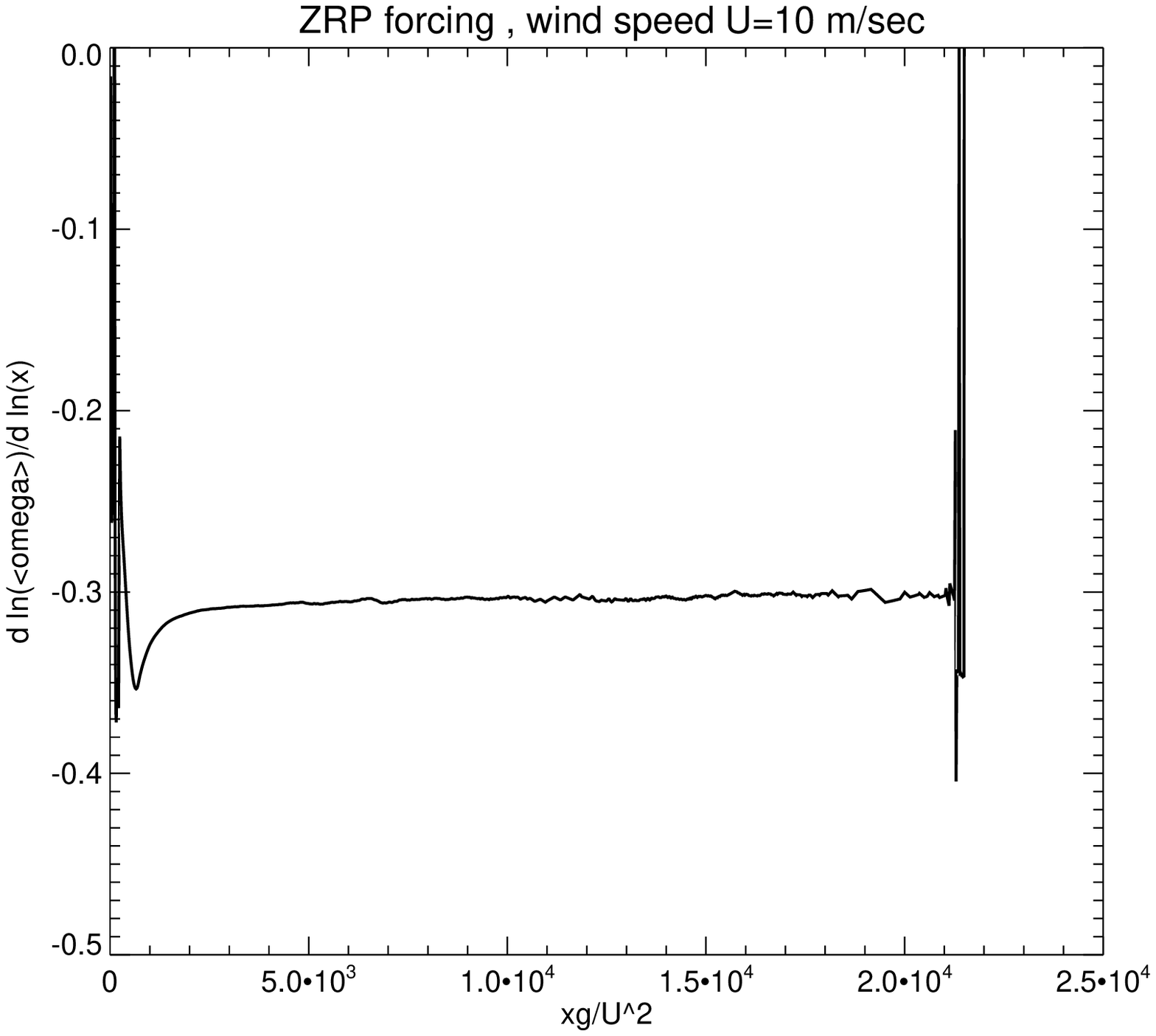}
                \caption{Exponent $q$ of the mean frequency dependence on fetch $x$}
                \label{MeanFreqB}
        \end{subfigure}
      \caption{}\label{MeanFreq}
\end{figure}

\noindent Dependence of mean frequency on the fetch, shown on Fig.\ref{MeanFreq}, also demonstrates perfect correspondence of numerical results and corresponding self-similar dependence Eq.(\ref{Qdep}) with $q=0.3$.


\begin{figure}
        \centering
        \begin{subfigure}[b]{0.45\textwidth}
                \includegraphics[width=\textwidth]{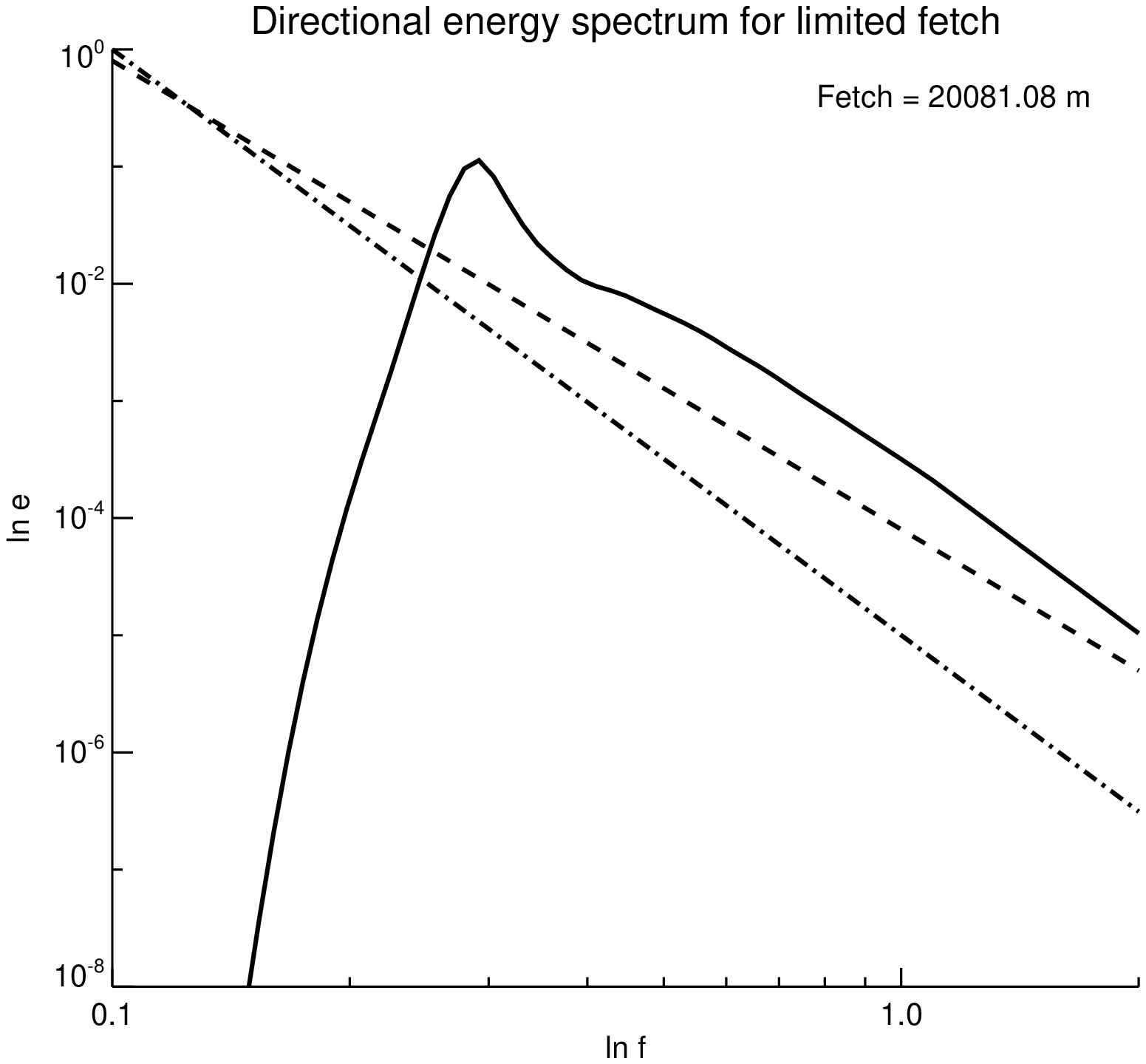}
                \caption{Logarithm of spectral energy density as a function of logarithm of frequency $f=\frac{\omega }{2\pi}$ - solid line. Dashed line - fit $\sim \omega ^{-4}$, dash-dotted line - fit $\sim \omega ^{-5}$.
}
                \label{SpecA}
        \end{subfigure}
\qquad 
        \begin{subfigure}[b]{0.45\textwidth}
                \includegraphics[width=\textwidth]{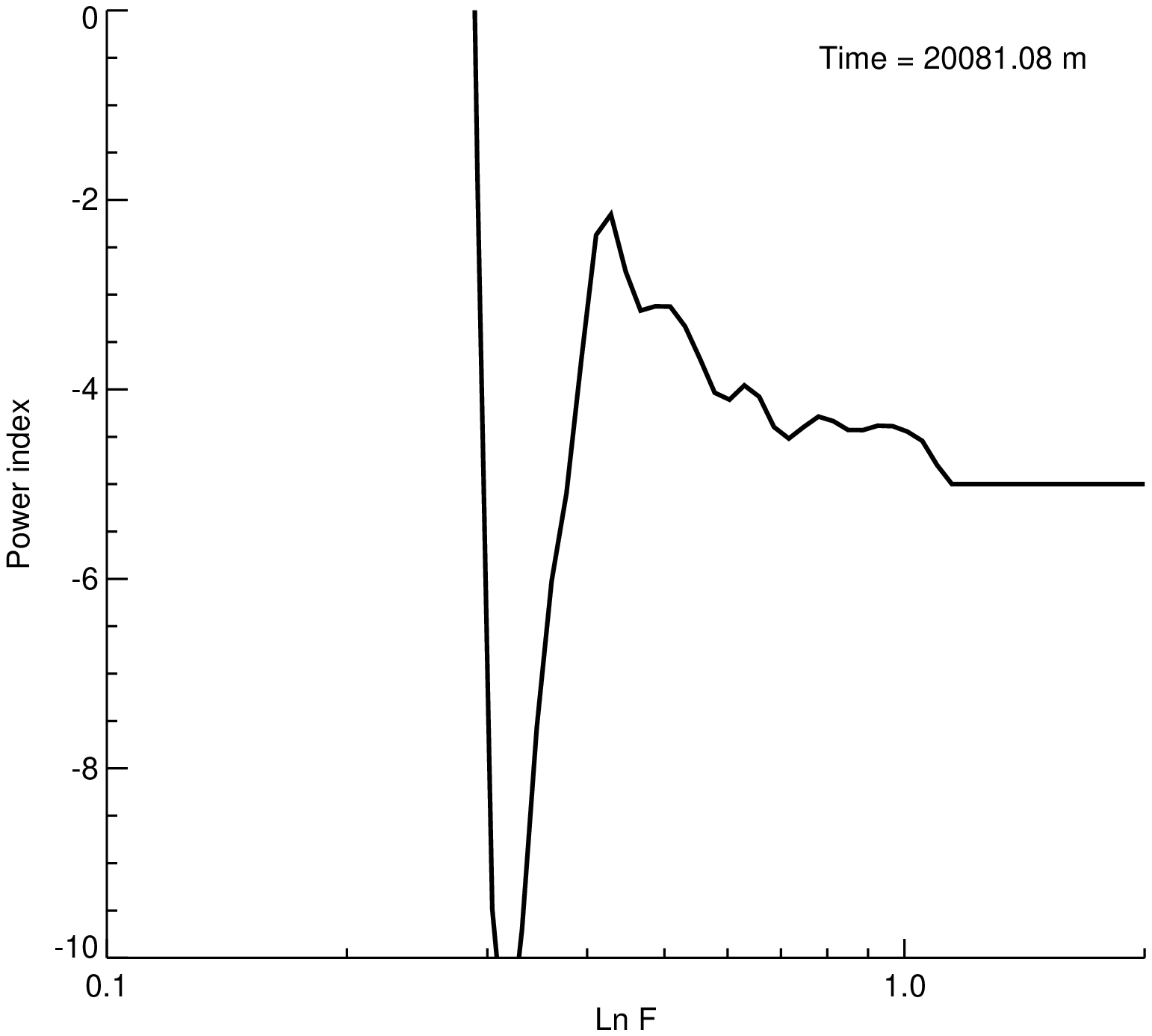}
                \caption{Local exponent of $\omega $ calculated from the solid line Fig.\ref{SpecA}.}
                \label{SpecB}
        \end{subfigure}
        \caption{}\label{Spec}
\end{figure}

\noindent Fig.\ref{SpecA} presents directional spectrum as a function of frequency in logarithmic coordinates. One can see that energy curve on the left figure consists of segments of:

\begin{enumerate}
\item  Spectral maximum area
\item  Kolmogorov-Zakharov  spectrum $\omega^{-4}$
\item  Phillips high frequency tail $\omega^{-5}$
\end{enumerate}

\noindent Fig.\ref{SpecB} shows log-log derivative of the spectral curve from  Fig.\ref{SpecA} figure, which corresponds to the exponent of the local power law. Again, one can see the areas corresponding to Kolmogorov-Zakharov index $-4$ and Phillips index $-5$. The value of the index to the left side from $-4$ plateau has the tendency to grow, which qualitatively corresponds to the ``inverse cascade'' Kolmogorov-Zakharov index $-11/3$.

\noindent One should stop on $\varepsilon \simeq \omega^{-4}$  asymptotics. It was observed in all our numerical experiments. It is Zakharov-Filonenko spectrum, which is the solution of equation

\begin{equation}
S_{nl} = 0
\end{equation}

\noindent It is predicted by weak-turbulent wind-wave turbulence theory and appears routinely in numerical experiments \cite{R2,R3,R4,R5,R6,R9}, see also \cite{R41,R42,R43,R44,R45}.

\noindent This spectrum is confirmed by multiple ocean field \cite{R47,R48,R49,R50}, wave tanks \cite{R51} and Lake George \cite{R29} measurements.

\noindent The "inverse cascade" spectrum $\varepsilon_\omega \simeq \omega^{-11/3}$ was also predicted by weak-turbulent theory \cite{R29}, \cite{R43,R44} and observed in numerical experiments \cite{R5}. Its field measurements, however, are less confident.

In reality, nonlinear $S_{nl}$ term is the leading term in the ocean energy balance \cite{R1}, \cite{R2}. It consists of two parts:
\begin{equation}
S_{nl} = F_k - \Gamma_k \varepsilon_k
\end{equation}
\noindent which almost compensate each other. Otherwise, one can not explain persistent presence of Zakharov -- Filonenko asymptotics $\varepsilon_\omega \simeq \omega^{-4}$.


\begin{figure}
        \centering
                \includegraphics[width=0.5\textwidth]{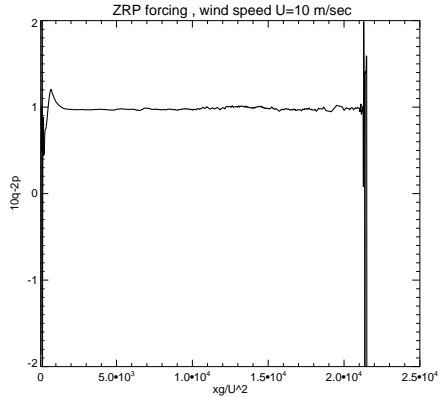}
                \caption{ Relation (10q-2p) as a function of the fetch $x$.}
	\label{MagicNumber}
\end{figure}

\noindent Fig.\ref{MagicNumber} presents relation "$(10q-2p)$ as a function of fetch $x$. It is in perfect accordance with self-similar prediction Eq.(\ref{MagicLaw}).


\noindent We conclude that alternative framework for $HE$ simulation reproduces the following analytical features of $HE$:

\begin{enumerate}
\item  Self-similar solutions with correct exponents
\item  Kolmogorov-Zakharov spectra $\sim \omega^{-4}$
\end{enumerate}

\noindent Table \ref{Table1} presents results of calculation of exponents $p$ and $q$ (see Eqs.(\ref{Pdep})-(\ref{Qdep})) for 14 different experimental observations with the last row corresponding to limited fetch growth numerical experiment within alternative framework. One can see good correspondence between theoretical, experimental and numerical values of $p$ and $q$.
\noindent 

\begin{table}
\centering
\begin{tabular}{|p{2.8in}|p{0.2in}|p{0.2in}|} \hline 
\textbf{Experiment} & $p$ & $q$ \\ \hline 
Babanin, Soloviev 1998 & 0.89 & 0.28 \\ \hline 
Walsh et al. (1989) US coast & 1.0 & 0.29 \\ \hline 
Kahma, Calkoen (1992) unstable & 0.94 & 0.28 \\ \hline 
Kahma, Pettersson (1994) & 0.93 & 0.28 \\ \hline 
JONSWAP by Davidan (1980)  & 1.0 & 0.28 \\ \hline 
JONSWAP by Phillips (1977)  & 1.0 & 0.25 \\ \hline 
Kahma, Calkoen (1992) composite  & 0.9 & 0.27 \\ \hline 
Kahma (1981, 1986) rapid growth  & 1.0 & 0.33 \\ \hline 
Kahma (1986) average growth  & 1.0 & 0.33 \\ \hline 
Donelan \textit{et al. }(1992) St Claire  & 1.0 & 0.33 \\ \hline 
Ross (1978), Atlantic, stable  & 1.1 & 0.27 \\ \hline 
Liu, Ross (1980), Lake Michigan, unstable  & 1.1 & 0.27 \\ \hline 
JONSWAP by Hasselmann et al. (1973)  & 1.0 & 0.33 \\ \hline 
Mitsuyasu et al. (1971)  & 1.0 & 0.33 \\ \hline 
ZRP numerics & 1.0 & 0.3 \\ \hline 
\end{tabular}
\caption{}
\label{Table1}
\end{table}

\section{Tests for separation of trustworthy wind input terms from non-physical ones}

\noindent As it was already discussed, there are plenty of historically developed parameterizations of wind input terms. Analysis of nonlinear properties of $HE$ in the form of specific self-similar solutions and Kolmogorov-Zakharov law for direct energy cascade allows us to propose the set of tests, which would allow separation of physically justified wind-input terms $S_{in}$ from non-physical ones. 

\noindent As such, we propose:

\begin{enumerate}
\item  Checking powers of observed energy and mean frequency dependencies along the fetch versus predicted by self-similar solutions.

\item  Checking the ``Magic relations'' Eq.(\ref{MagicLaw}) between exponents $p$ and $q$ for observed energy and frequency dependencies along the fetch.

\item  Checking exponents of directional spectral energy dependencies versus Kolmogorov-Zakharov exponent $-4$.
\end{enumerate}

\noindent We applied such tests to the results of $HE$ simulations which used the following popular wind input terms within alternative framework:

\noindent 

\begin{enumerate}

\item  Chalikov $S_{in}$ term \cite{R25,R15}

\item  Snyder $S_{in}$ term \cite{R30}

\item   Hsiao-Shemdin $S_{in}$ term \cite{R31}

\item  $WAM3$ $S_{in}$ term \cite{R52}

\end{enumerate}

\section{Test of Chalikov wind input term}

\begin{figure}
        \centering
        \begin{subfigure}[b]{0.45\textwidth}
                \includegraphics[width=\textwidth]{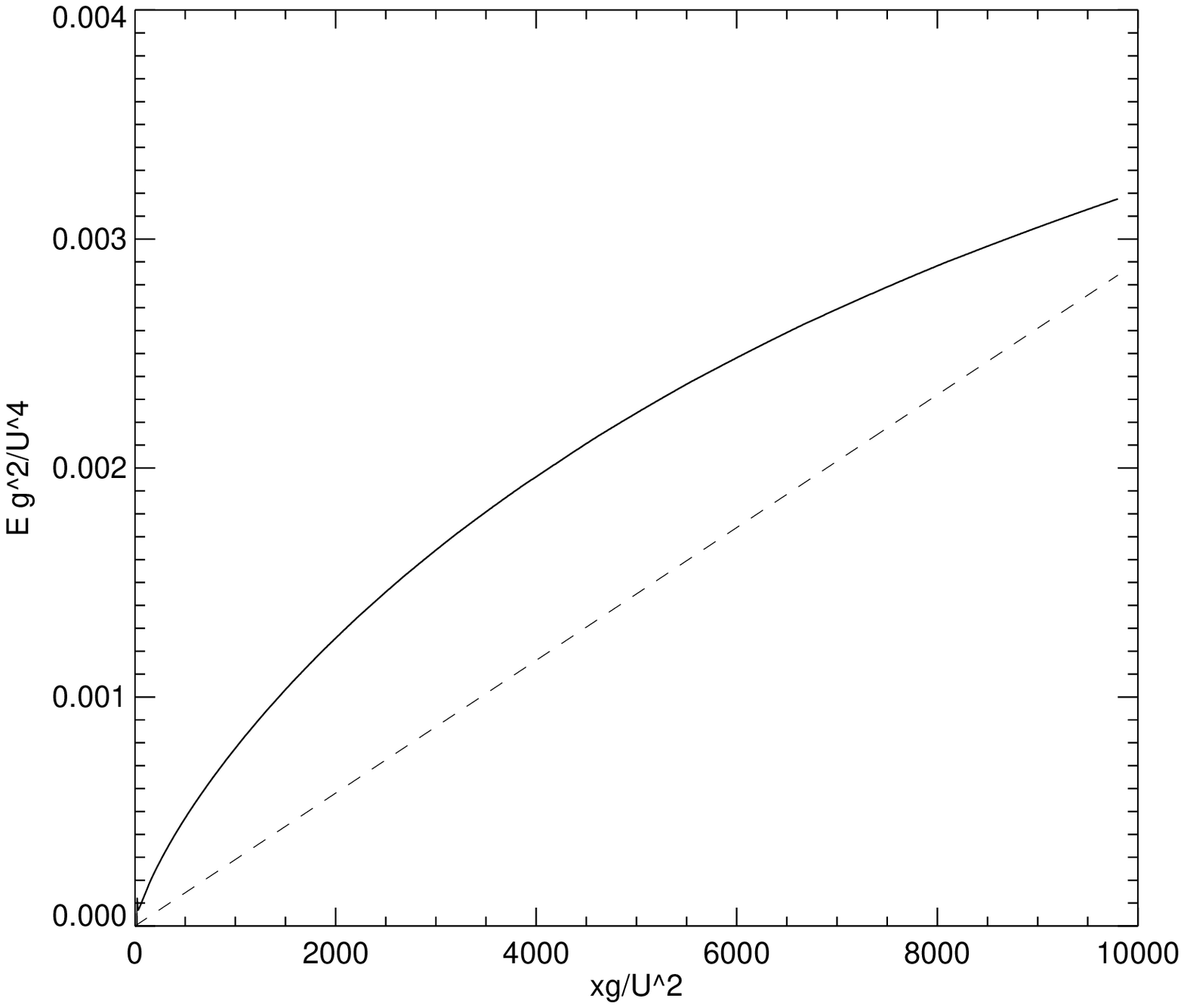}
                \caption{}
                \label{ChalikovFormA}
        \end{subfigure}
\qquad
        \begin{subfigure}[b]{0.45\textwidth}
                \includegraphics[width=\textwidth]{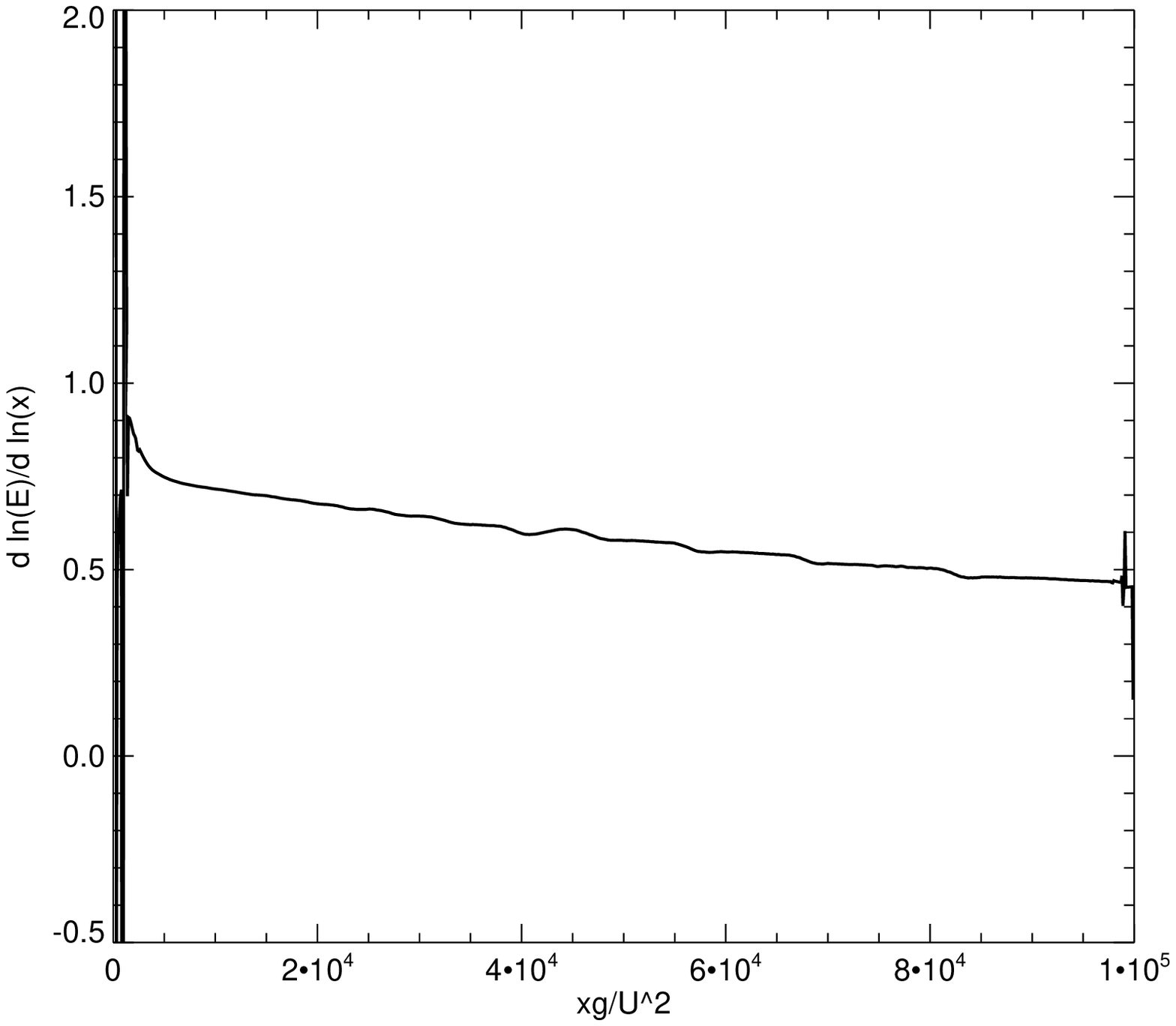}
                \caption{}
                \label{ChalikovFormB}
        \end{subfigure}
	\caption{Same as Fig.\ref{AltForm}, but for Chalikov $S_{in}$}
	\label{ChalikovForm}
\end{figure}

\noindent Fig.\ref{ChalikovForm} shows that total energy growth along the fetch significantly exceeds observed in $ZRP$ simulation, and value of the corresponding exponent significantly deviates from theoretical value $p=1.0$


\begin{figure}
        \centering
        \begin{subfigure}[b]{0.45\textwidth}
                \includegraphics[width=\textwidth]{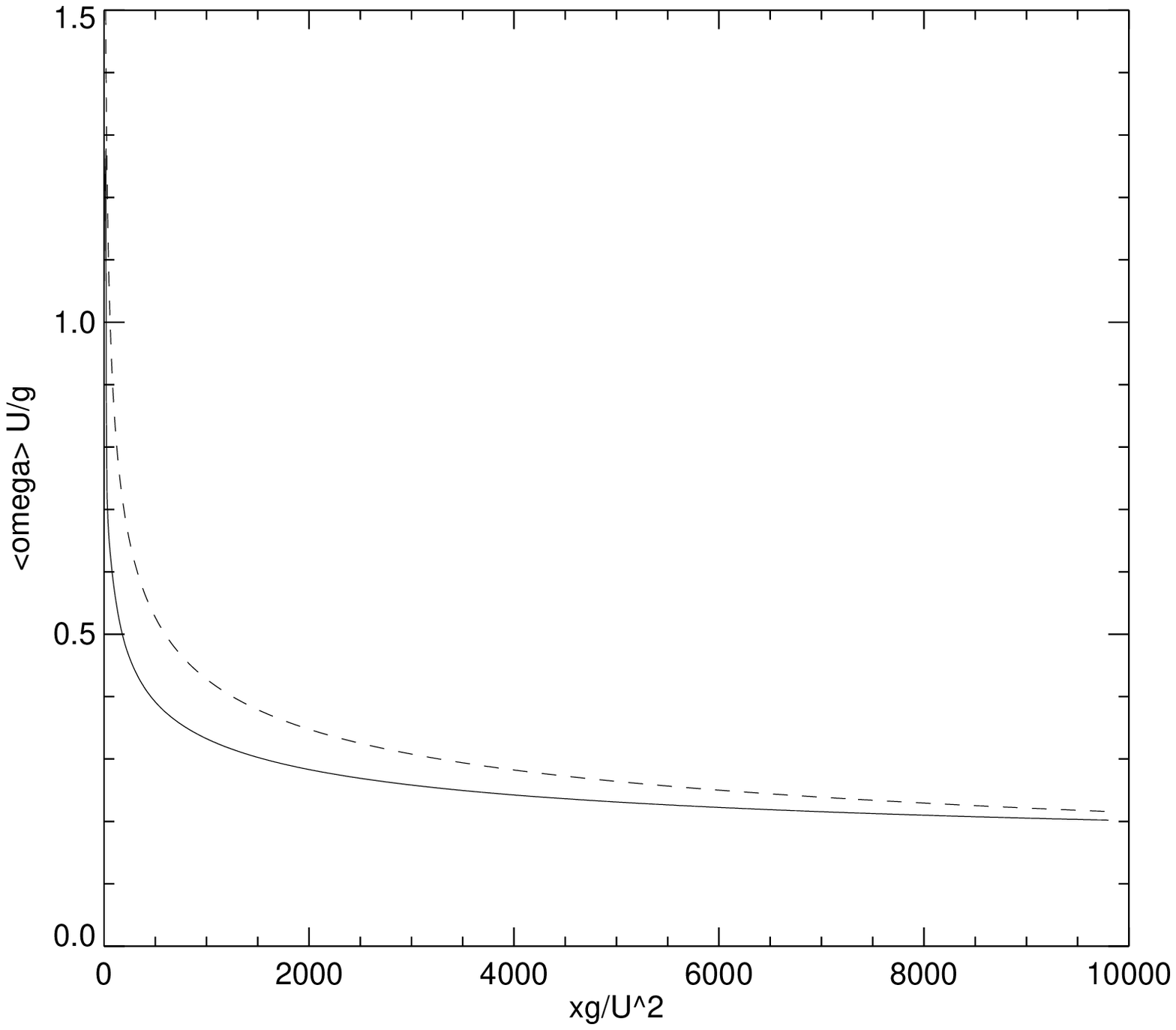}
                \caption{}
                \label{MeanFreqChalikovA}
        \end{subfigure}
\qquad
        \begin{subfigure}[b]{0.45\textwidth}
                \includegraphics[width=\textwidth]{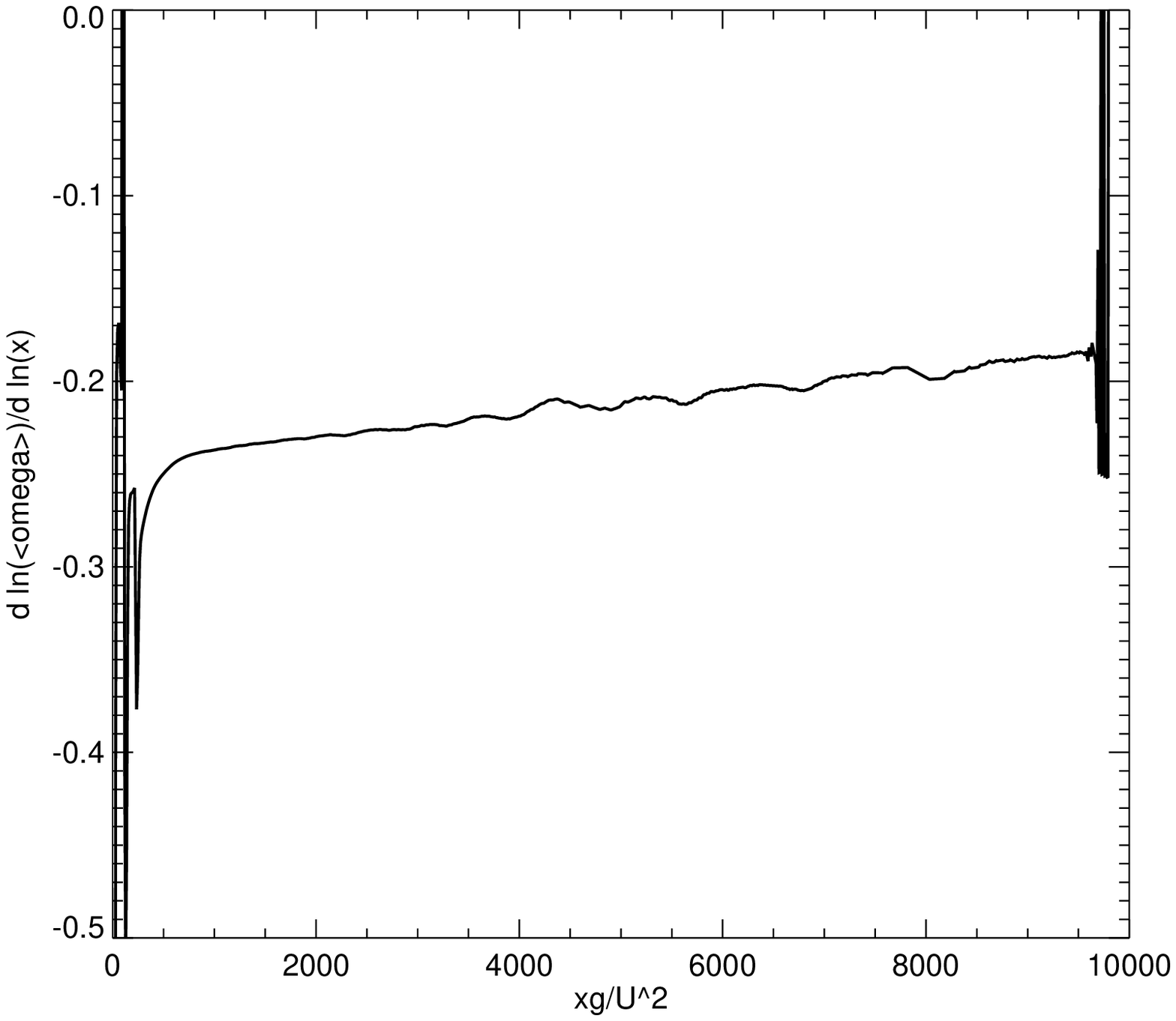}
                \caption{}
                \label{MeanFreqChalikovB}
        \end{subfigure}
	\caption{ Same as Fig.\ref{MeanFreq}, but for Chalikov $S_{in}$ }
	\label{MeanFreqChalikov}
\end{figure}

\noindent Dependence of mean frequency against the fetch, shown on Fig.\ref{MeanFreqChalikov}, also deviates from $ZRP$ numerical results and corresponding self-similar exponent $q=0.3$


\begin{figure}
        \centering
        \begin{subfigure}[b]{0.45\textwidth}
                \includegraphics[width=\textwidth]{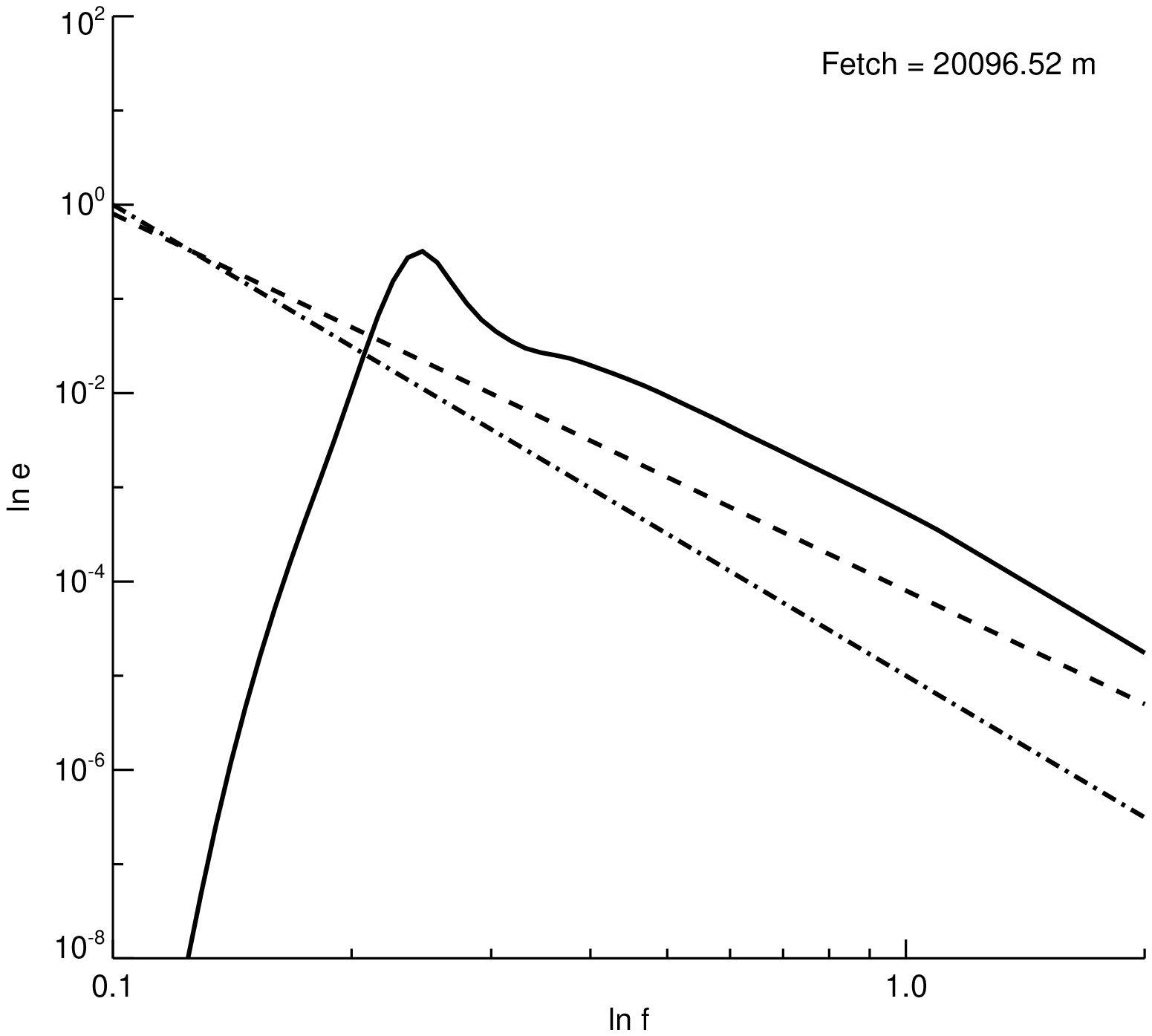}
                \caption{}
                \label{SpecChalikovA}
        \end{subfigure}
\qquad 
        \begin{subfigure}[b]{0.45\textwidth}
                \includegraphics[width=\textwidth]{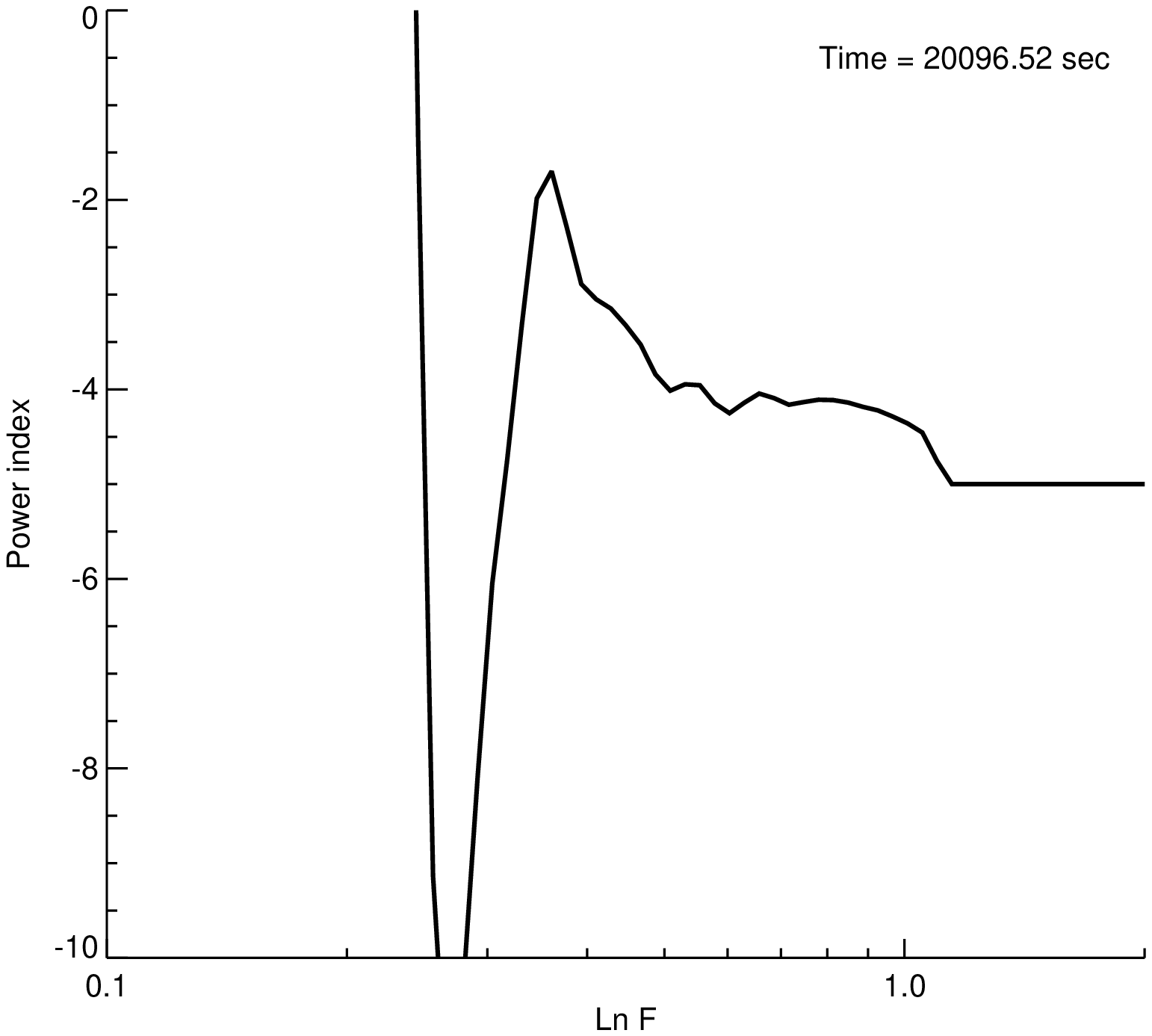}
                \caption{}
                \label{SpecChalikovB}
        \end{subfigure}
        \caption{Same as Fig.3, but for Chalikov $S_{in}$}
	\label{SpecChalikov}
\end{figure}

\noindent Left side of Fig.\ref{SpecChalikovA} presents directional spectrum as a function of frequency in logarithmic coordinates. One can see that similar to $ZRP$ case we observe:

\noindent 

\begin{enumerate}
\item  Spectral maximum area

\item  Kolmogorov-Zakharov segment $\sim \omega ^{-4}$

\item  Phillips high frequency tail $\sim \omega^{-5}$
\end{enumerate}

\noindent Fig.\ref{SpecChalikovB} shows log-log derivative of the energy curve from  Fig.\ref{SpecChalikovA}, corresponding to the exponent of the local power law. Again, one can see the areas corresponding to Kolmogorov-Zakharov index $-4$ and Phillips index $-5$. The value of the index to the left side of $-4$  plateau has a tendency to grow, which qualitatively corresponds to the  Kolmogorov-Zakharov inverse cascade index $-11/3$.


\noindent 

\begin{figure}
        \centering
	\includegraphics[width=0.5\textwidth]{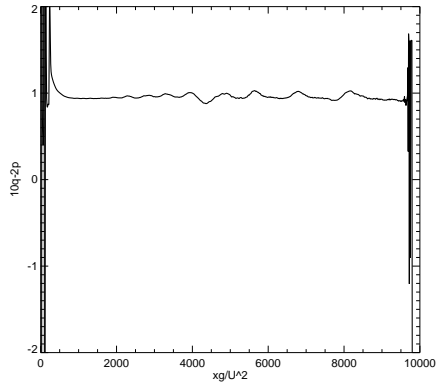}
	\caption{"Magic number" $10q-2p$ as a function of the fetch $x$ for Chalikov wind input term.}
	\label{MagicNumberChalikov}
\end{figure}

\noindent Fig.\ref{MagicNumberChalikov} presents combination $(10q-2p)$ as a function of fetch distance $x$. It is surprising that it is in perfect accordance with the relation Eq.(\ref{MagicLaw}). It mean that despite incorrect values $p$ and $q$ along the fetch, their combination $(10q-2p)$ still holds in complete accordance with theoretical prediction, i.e. self-similarity is fulfilled locally.


\section{Test of Snyder wind input term}

\begin{figure}
        \centering
        \begin{subfigure}[b]{0.45\textwidth}
                \includegraphics[width=\textwidth]{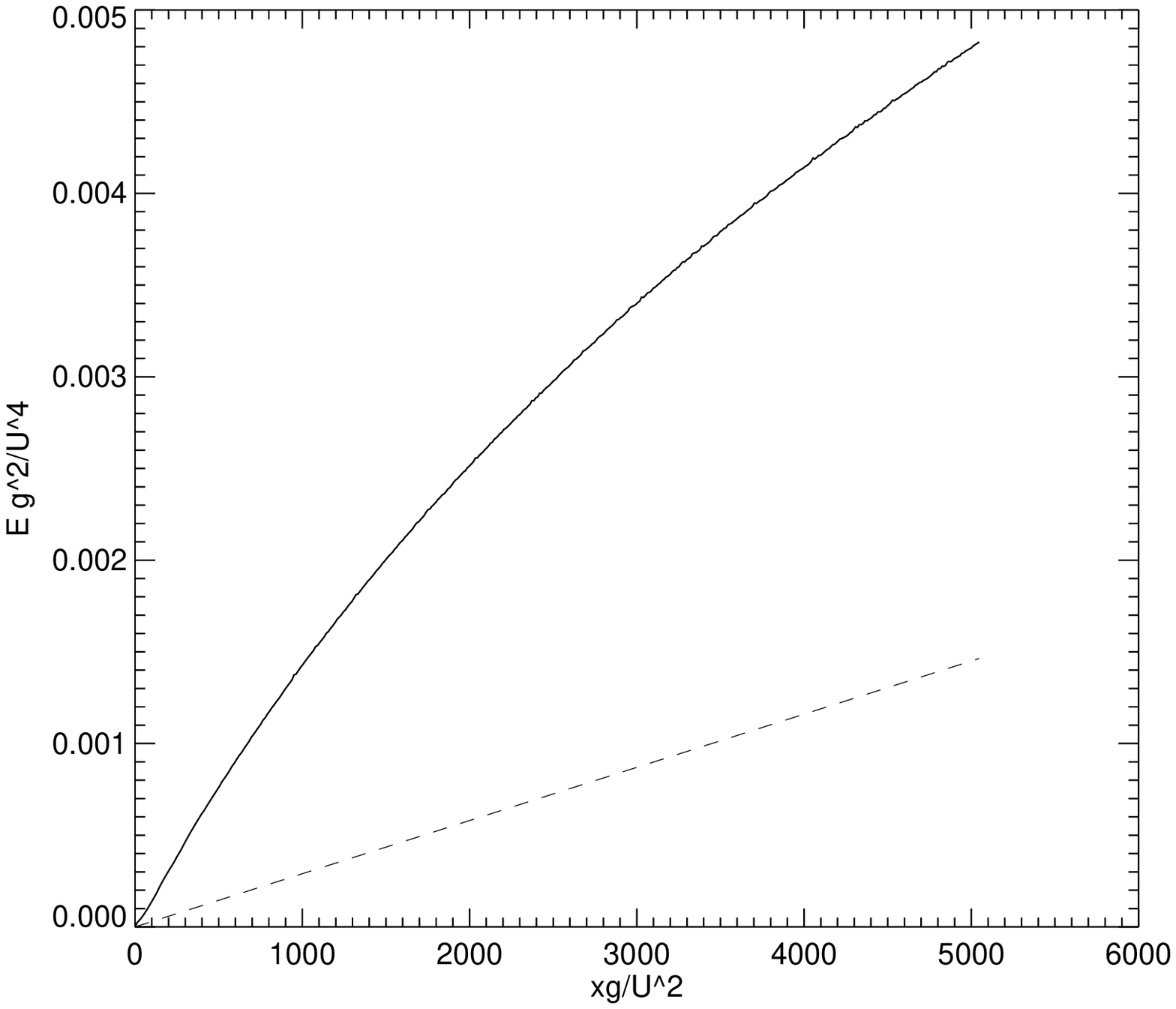}
                \caption{}
                \label{SnyderFormA}
        \end{subfigure}
\qquad
        \begin{subfigure}[b]{0.45\textwidth}
                \includegraphics[width=\textwidth]{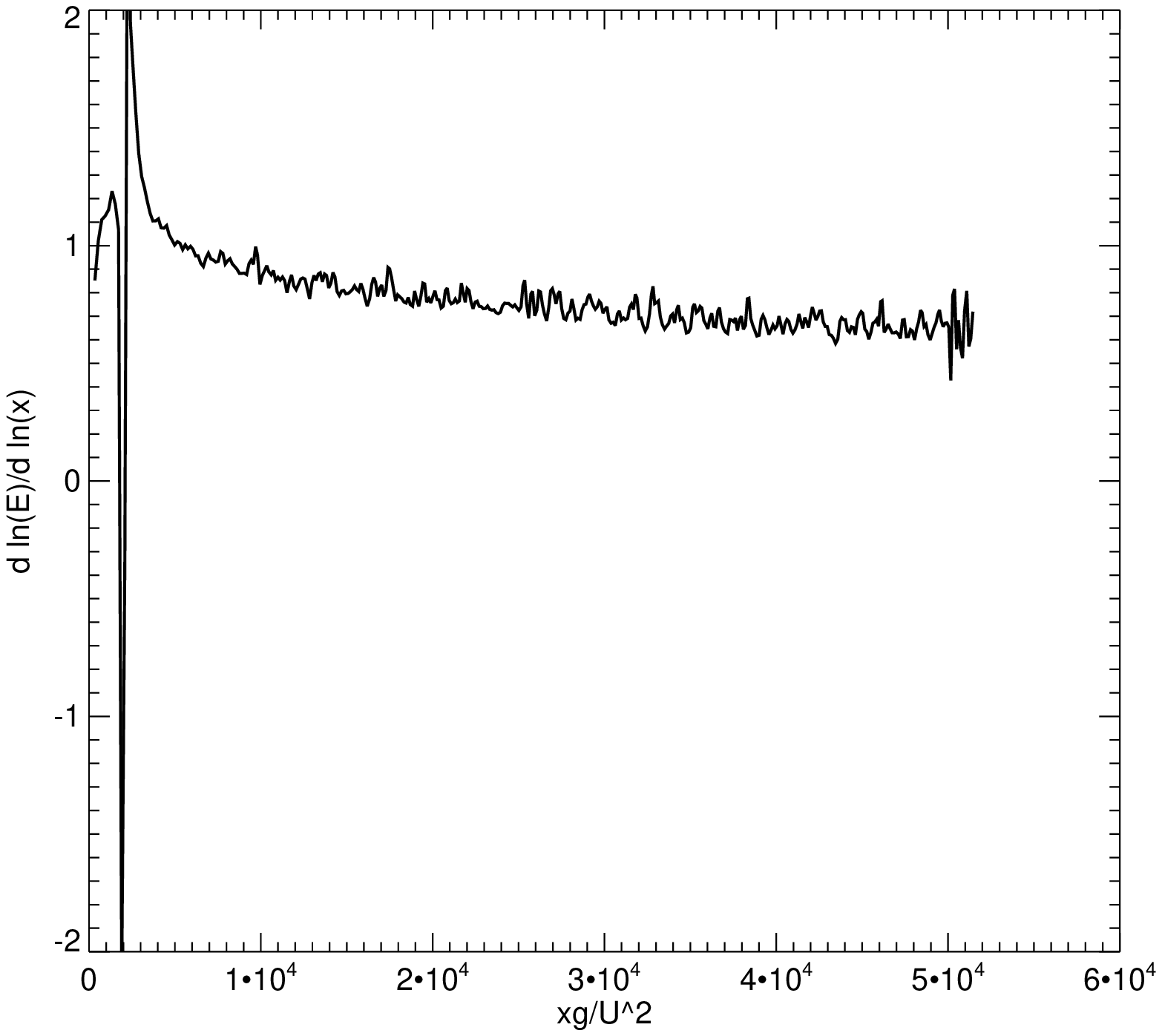}
                \caption{}
                \label{SnyderFormB}
        \end{subfigure}
	\caption{Same as Fig.\ref{AltForm}, but for Snyder $S_{in}$}
	\label{SnyderForm}
\end{figure}

\noindent Fig.\ref{SnyderForm} shows that total energy growth along the fetch significantly exceeds $ZRP$ case, but has the value of growth exponent close to $p=1.0$ versus fetch coordinate $x$. 


\begin{figure}
        \centering
        \begin{subfigure}[b]{0.45\textwidth}
                \includegraphics[width=\textwidth]{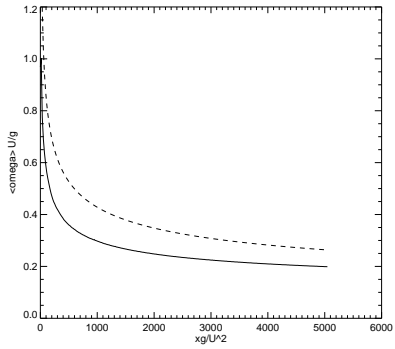}
                \caption{}
                \label{MeanFreqSnyderA}
        \end{subfigure}
\qquad
        \begin{subfigure}[b]{0.45\textwidth}
                \includegraphics[width=\textwidth]{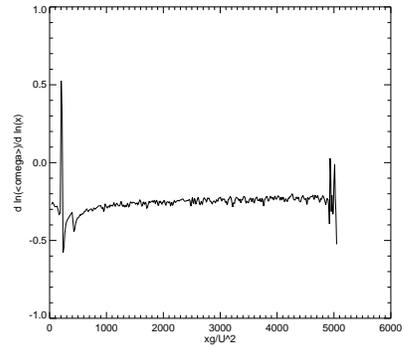}
                \caption{}
                \label{MeanFreqSnyderB}
        \end{subfigure}
      \caption{Same as Fig.\ref{MeanFreq}, but for Snyder $S_{in}$}\label{MeanFreqSnyder}
\end{figure}

\noindent Dependence of mean frequency against the fetch shown on Fig.\ref{MeanFreqSnyder} is lower than $ZRP$ numerical results, but has fairly close value to self-similar solution index $q=0.3$.


\begin{figure}
        \centering
        \begin{subfigure}[b]{0.45\textwidth}
                \includegraphics[width=\textwidth]{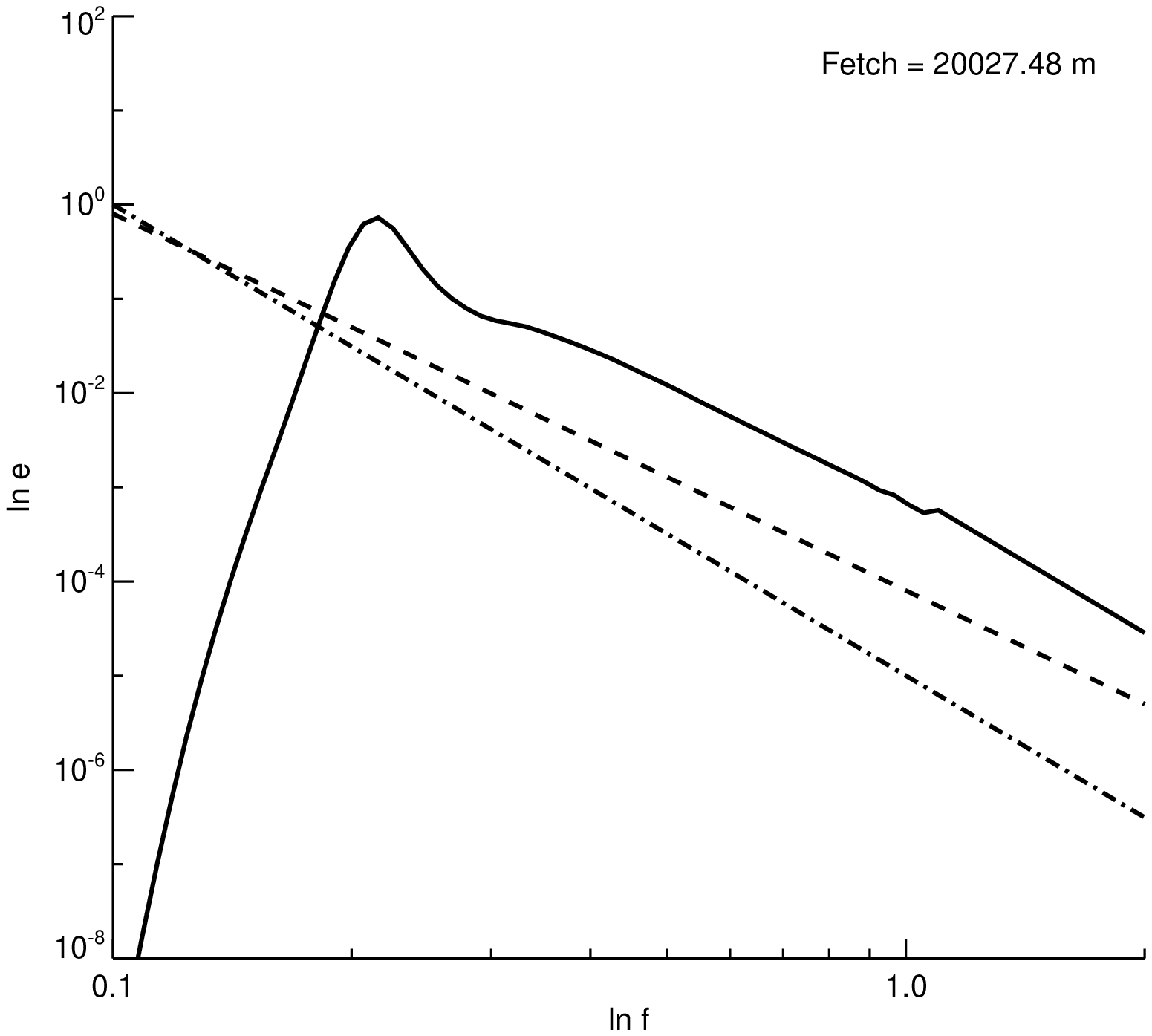}
                \caption{}
                \label{SpecSnyderA}
        \end{subfigure}
\qquad 
        \begin{subfigure}[b]{0.45\textwidth}
                \includegraphics[width=\textwidth]{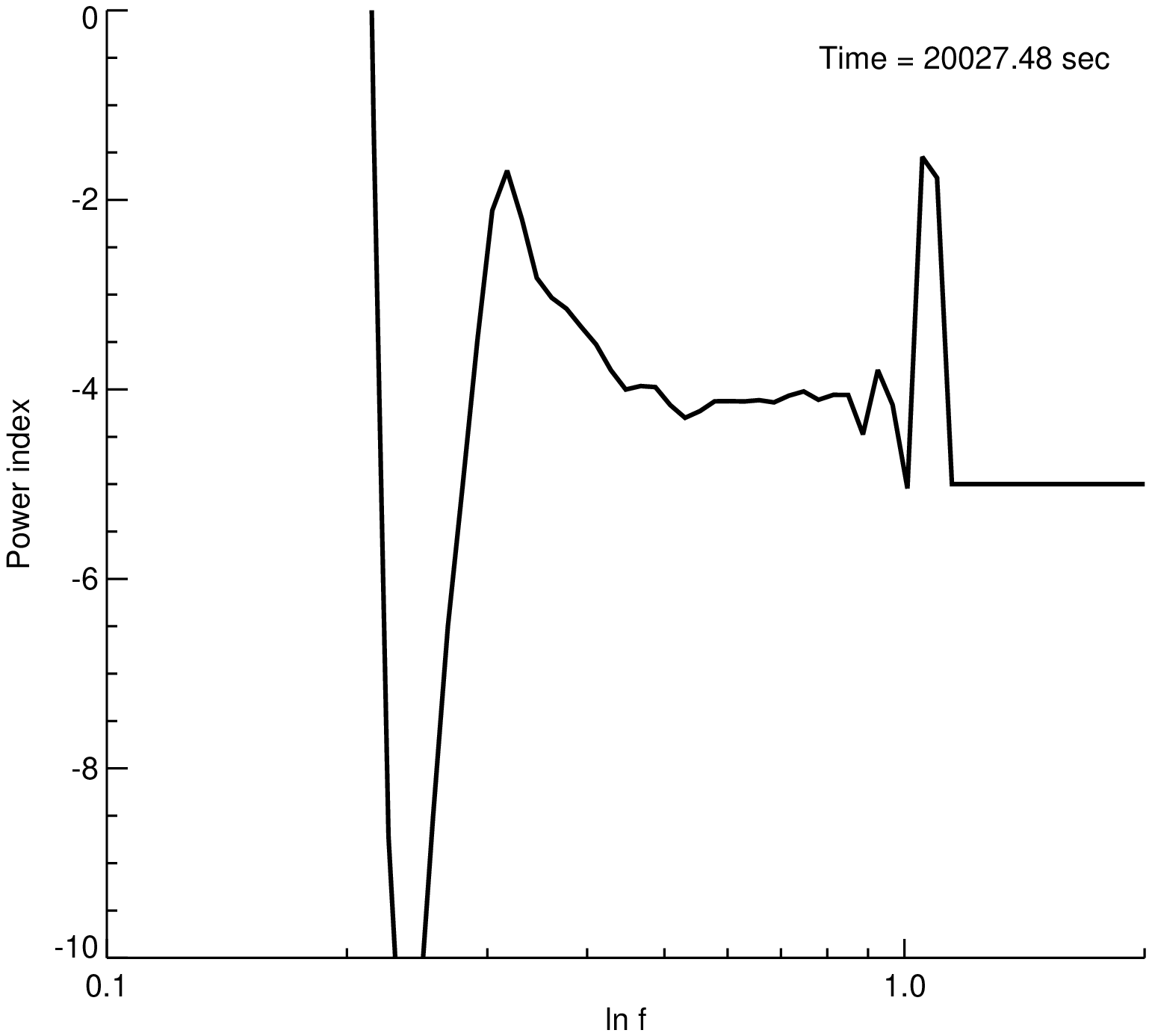}
                \caption{}
                \label{SpecSnyderB}
        \end{subfigure}
        \caption{Same as Fig.\ref{Spec} but for Snyder $S_{in}$}
	\label{SpecSnyder}
\end{figure}

Fig.\ref{SpecSnyderA} presents directional spectrum as a function of frequency in logarithmic coordinates. One can see:

\begin{enumerate}
\item  Spectral maximum area

\item  Kolmogorov-Zakharov segment $\sim \omega ^{-4}$

\item  Phillips high frequency tail $\sim \omega ^{-5}$
\end{enumerate}

\noindent Fig.\ref{SpecSnyderB} shows log-log derivative of the energy curve from Fig.\ref{SpecSnyderA}, which corresponds to the exponent of the local power law. Again, one can see the areas corresponding to Kolmogorov-Zakharov index $-4$ and Phillips index $-5$. The value of the index to the left side of $-4$ has a tendency to grow, which qualitatively corresponds to the ``inverse cascade'' Kolmogorov-Zakharov index $-11/3$.


\begin{figure}
        \centering
	\includegraphics[width=0.5\textwidth]{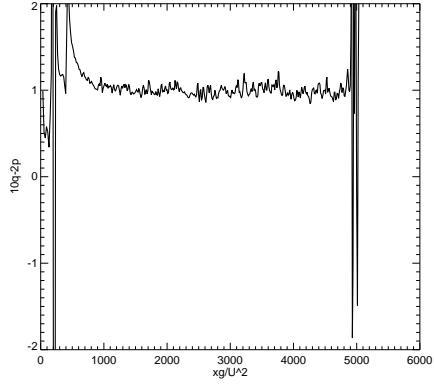}
	\caption{Relation $(10q-2p)$ as a function of the fetch $x$ for Snyder wind input term.}
	\label{MagicNumberSnyder}
\end{figure}

\noindent Fig.\ref{MagicNumberSnyder} presents the combination $(10q-2p)$ as the function of the fetch. Again, it is in perfect accordance with the theoretical relation Eq.(\ref{MagicLaw}). As in Chalikov case it means that despite not perfect values of $p$ and $q$ and wrong energy growth along the fetch, their combination $(10q-2p)$ still holds in complete accordance with theoretical prediction, i.e. self-similarity is also fulfilled locally in Snyder case.


\section{Test of \textit{Hsiao-Shemdin} wind input term}

\begin{figure}
        \centering
        \begin{subfigure}[b]{0.45\textwidth}
                \includegraphics[width=\textwidth]{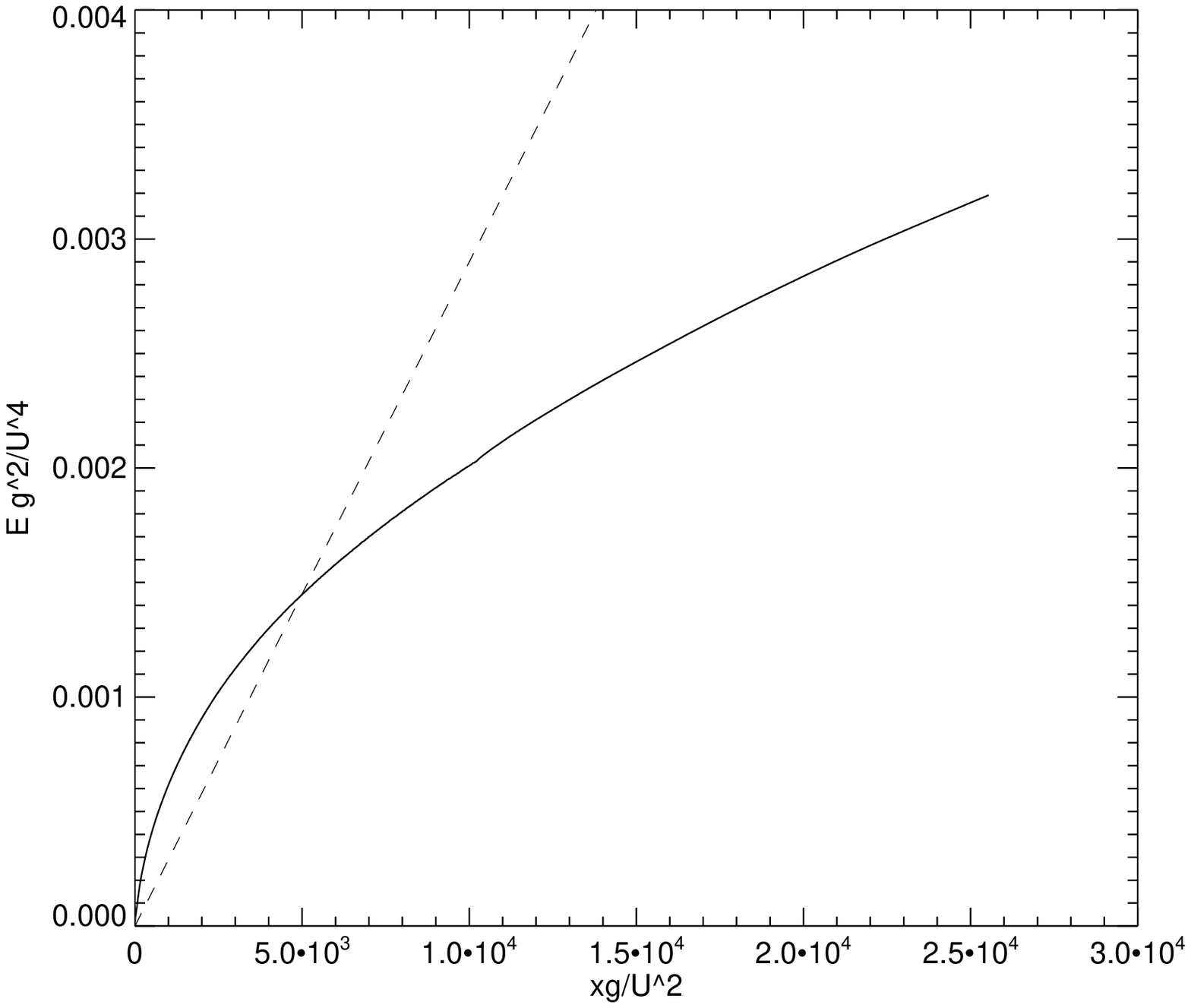}
                \caption{}
                \label{HSFormA}
        \end{subfigure}
\qquad
        \begin{subfigure}[b]{0.45\textwidth}
                \includegraphics[width=\textwidth]{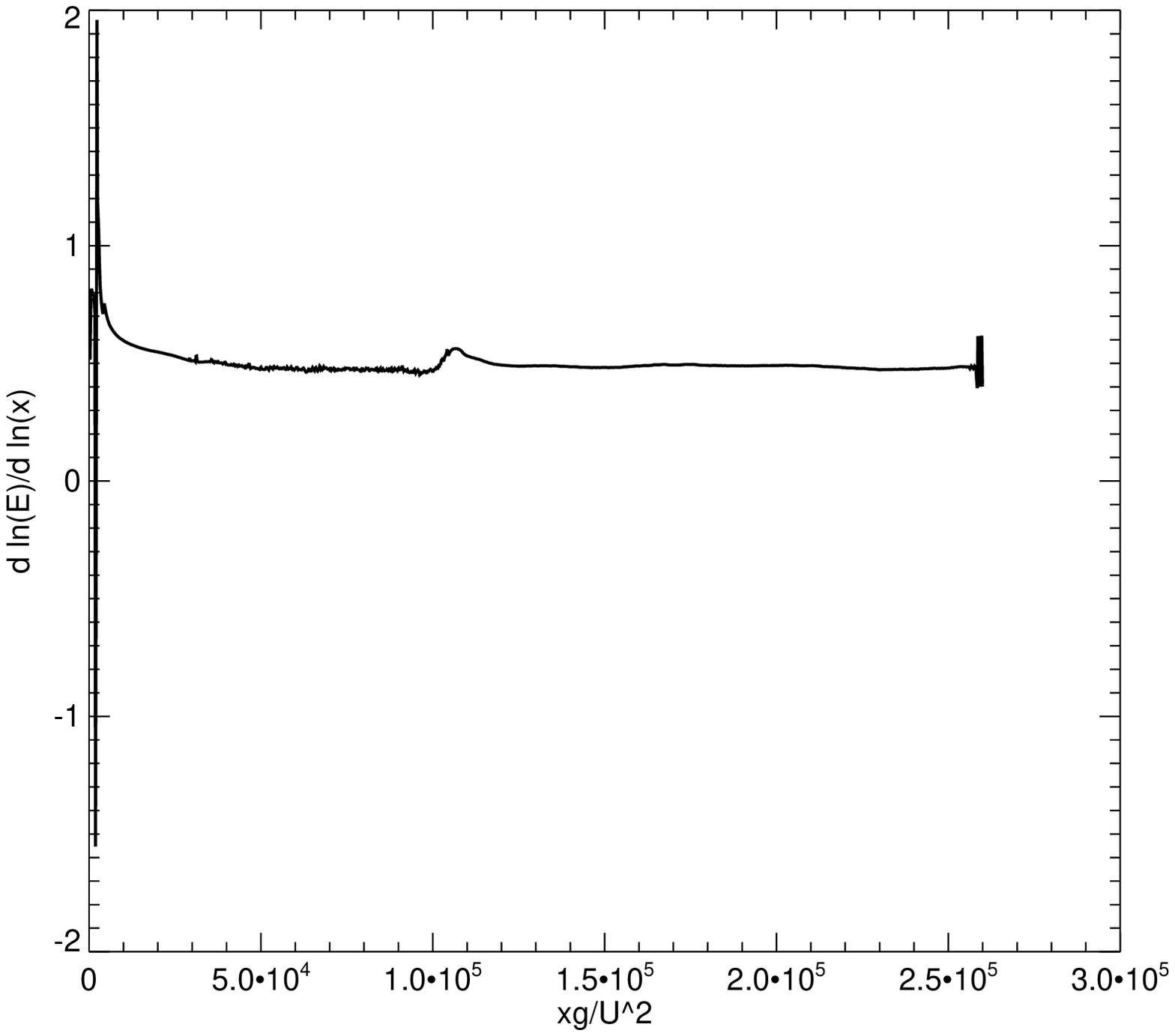}
                \caption{}
                \label{HSFormB}
        \end{subfigure}
	\caption{Same as Fig.\ref{AltForm}, but for $Hsiao-Shemdin$ $S_{in}$}
	\label{HSForm}
\end{figure}

\noindent Fig.\ref{HSForm} shows that total energy growth along the fetch strongly underestimates $ZRP$ simulation, and has the asymptotic value of exponent $p \approx 0.5$.


\begin{figure}
        \centering
        \begin{subfigure}[b]{0.45\textwidth}
                \includegraphics[width=\textwidth]{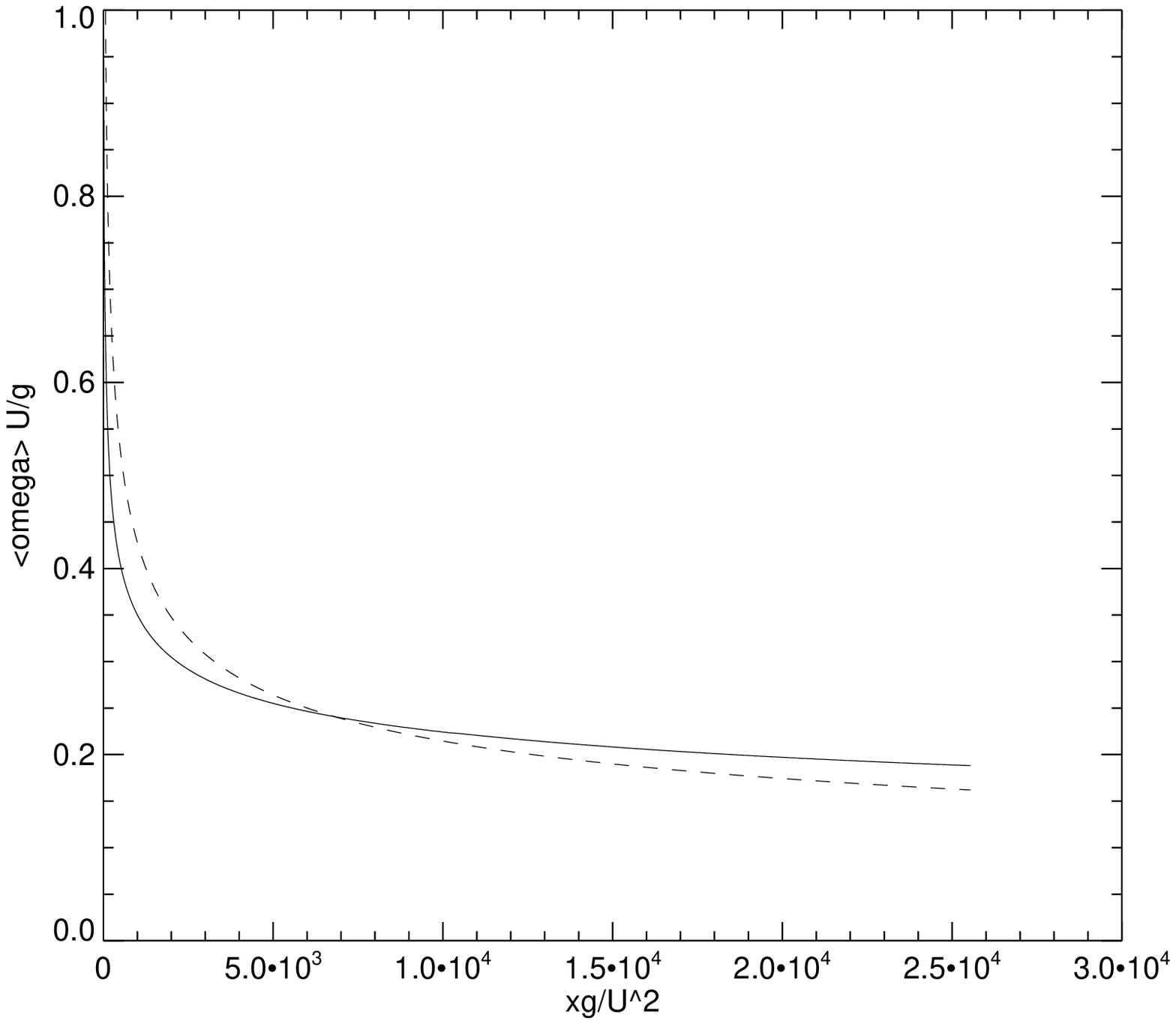}
                \caption{}
                \label{MeanFreqHSA}
        \end{subfigure}
\qquad
        \begin{subfigure}[b]{0.45\textwidth}
                \includegraphics[width=\textwidth]{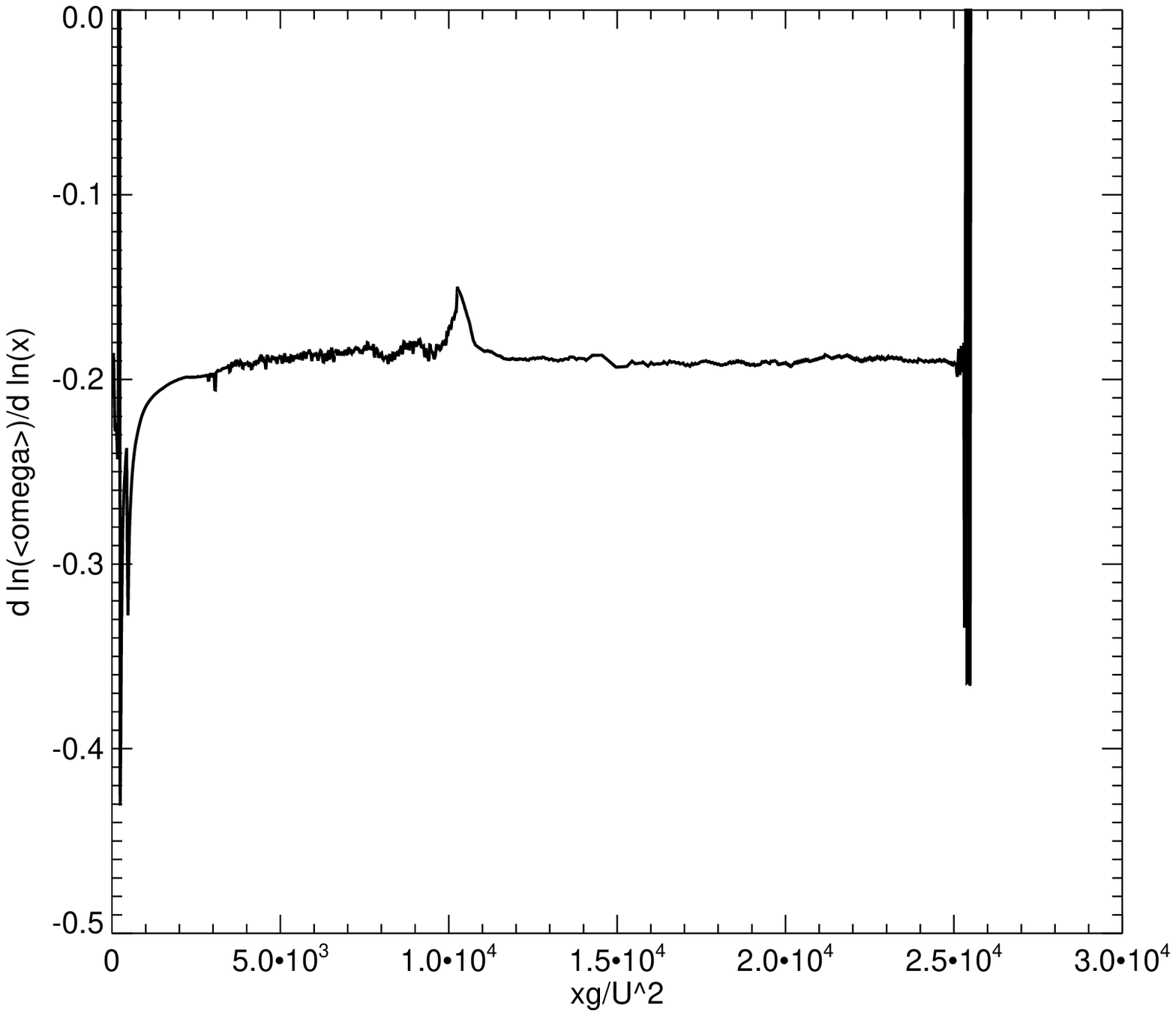}
                \caption{}
                \label{MeanFreqHSB}
	\end{subfigure}
	\caption{Same as Fig.\ref{MeanFreq}, but for $Hsiao-Shemdin$ $S_{in}$}
	\label{MeanFreqHS}
\end{figure}

\noindent Dependence of the mean frequency against the fetch shown on Fig.\ref{MeanFreqHS} demonstrates discrepancy with $ZRP$ results and asymptotic value of index $q \approx 0.21$.


\begin{figure}
        \centering
        \begin{subfigure}[b]{0.45\textwidth}
                \includegraphics[width=\textwidth]{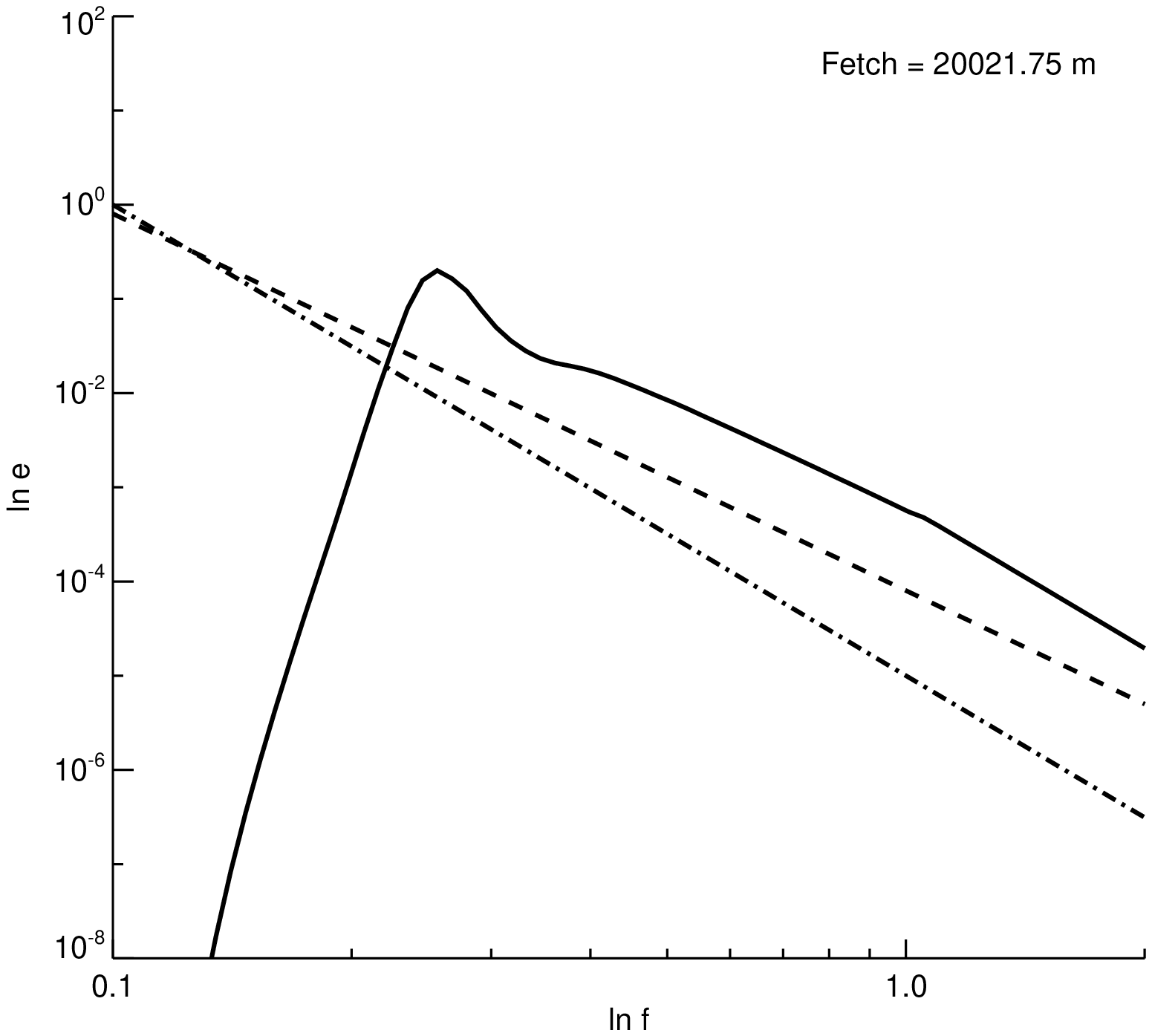}
                \caption{}
                \label{SpecHSA}
        \end{subfigure}
\qquad 
        \begin{subfigure}[b]{0.45\textwidth}
                \includegraphics[width=\textwidth]{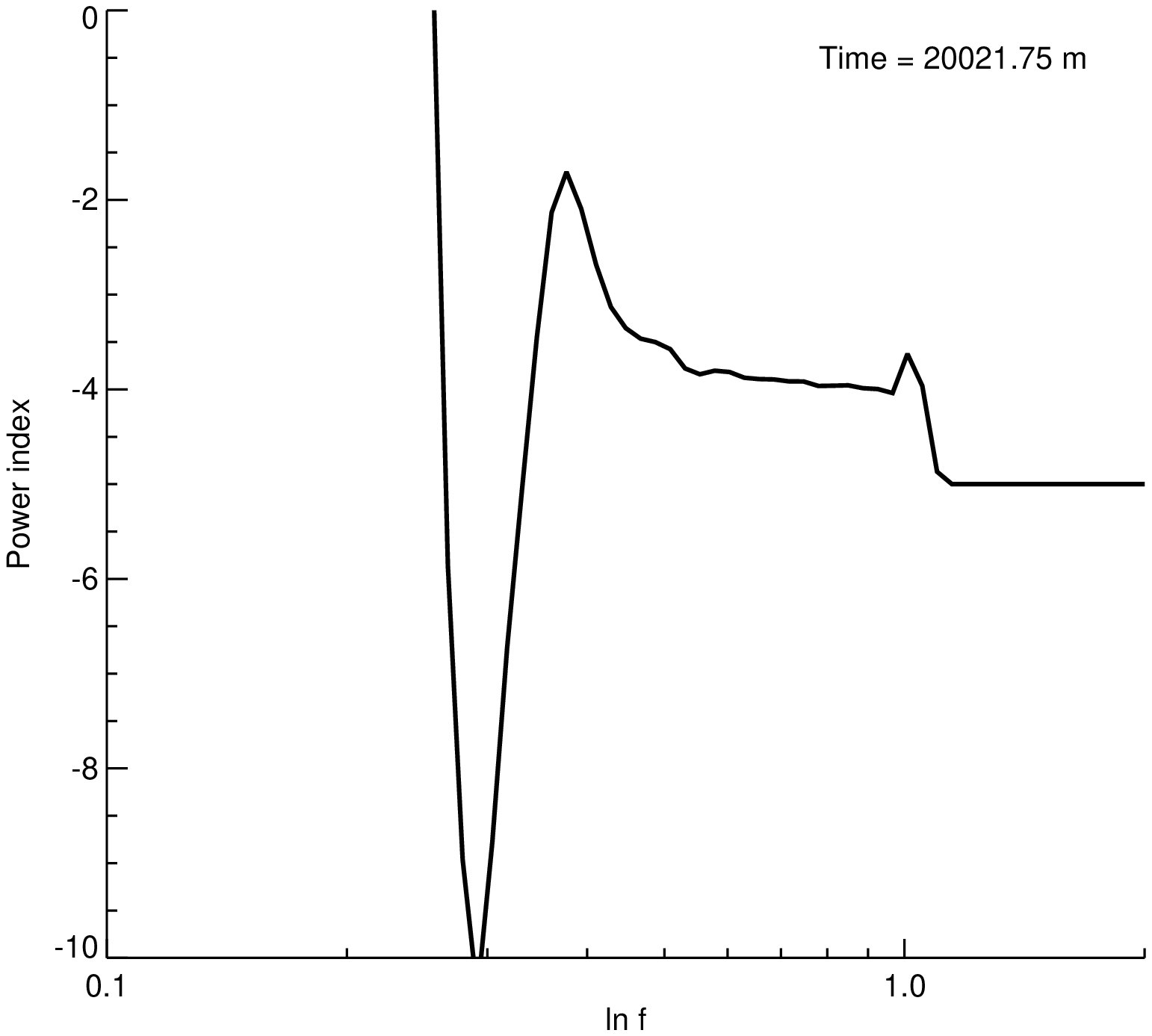}
                \caption{}
                \label{SpecHSB}
        \end{subfigure}
        \caption{Same as Fig.\ref{Spec} but for $Hsio-Shemdin$ $S_{in}$}
	\label{SpecHS}
\end{figure}

\noindent Fig.\ref{SpecHS} presents directional spectrum as a function of frequency in logarithmic coordinates. One can see:

\begin{enumerate}

\item Spectral maximum area

\item  Kolmogorov-Zakharov segment $\sim \omega^{-4}$

\item  Phillips high frequency tail $\sim \omega^{-5}$
\end{enumerate}

\noindent Fig.\ref{SpecHSB} shows log-log derivative of the energy curve from Fig.\ref{SpecHSA}, corresponding to the exponent of the local power law. Again, one can see the areas corresponding  Kolmogorov-Zakharov index $-4$ and Phillips index $-5$. The value of the index to the left side of $-4$ plateau has the tendency to grow, which qualitatively corresponds to the ``inverse cascade'' Kolmogorov-Zakharov index $-11/3$.


\begin{figure}
        \centering
	\includegraphics[width=0.5\textwidth]{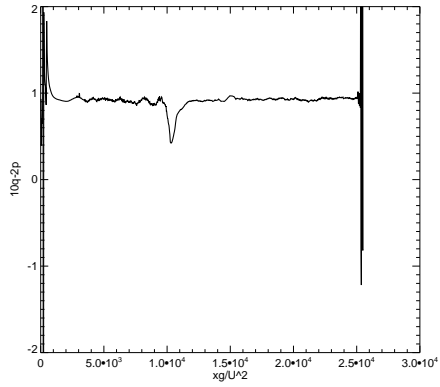}
	\caption{ Relation $(10q-2p)$ as a function of the fetch $x$ for $Hsiao-Shemdin$ wind input term.}
	\label{MagicNumberHS}
\end{figure}

\noindent Fig.\ref{MagicNumberHS} presents combination $(10q-2p)$ as the function of the fetch coordinate $x$. It is in total agreement with the theoretical predictions Eq.(\ref{MagicLaw}), which means that self-similarity is fulfilled locally in $Hsiao-Shemdin$ case.


\section{Test of \textit{WAM3} wind input term}

\begin{figure}
        \centering
        \begin{subfigure}[b]{0.45\textwidth}
                \includegraphics[width=\textwidth]{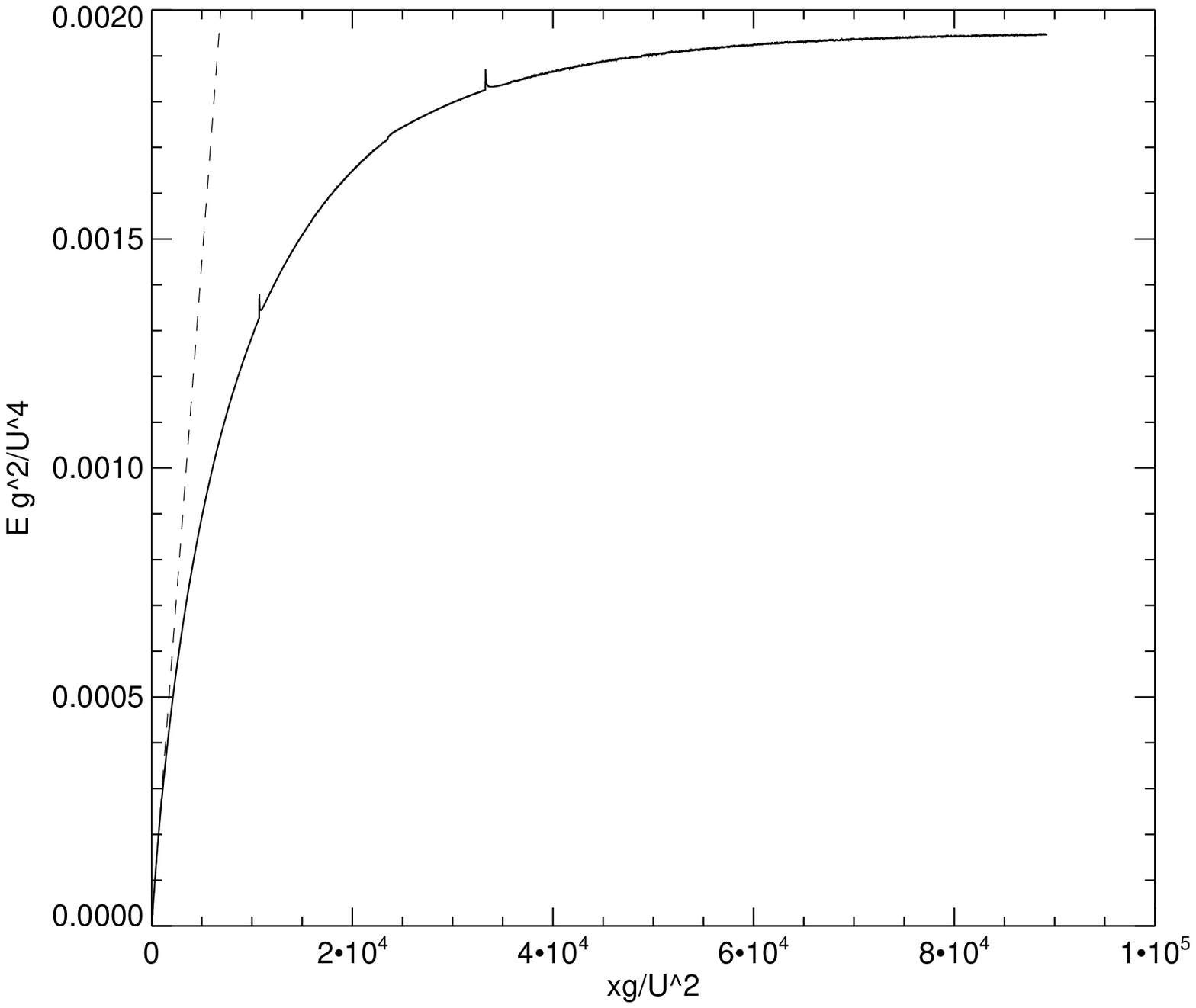}
                \caption{}
                \label{WAM3FormA}
        \end{subfigure}
\qquad
        \begin{subfigure}[b]{0.45\textwidth}
                \includegraphics[width=\textwidth]{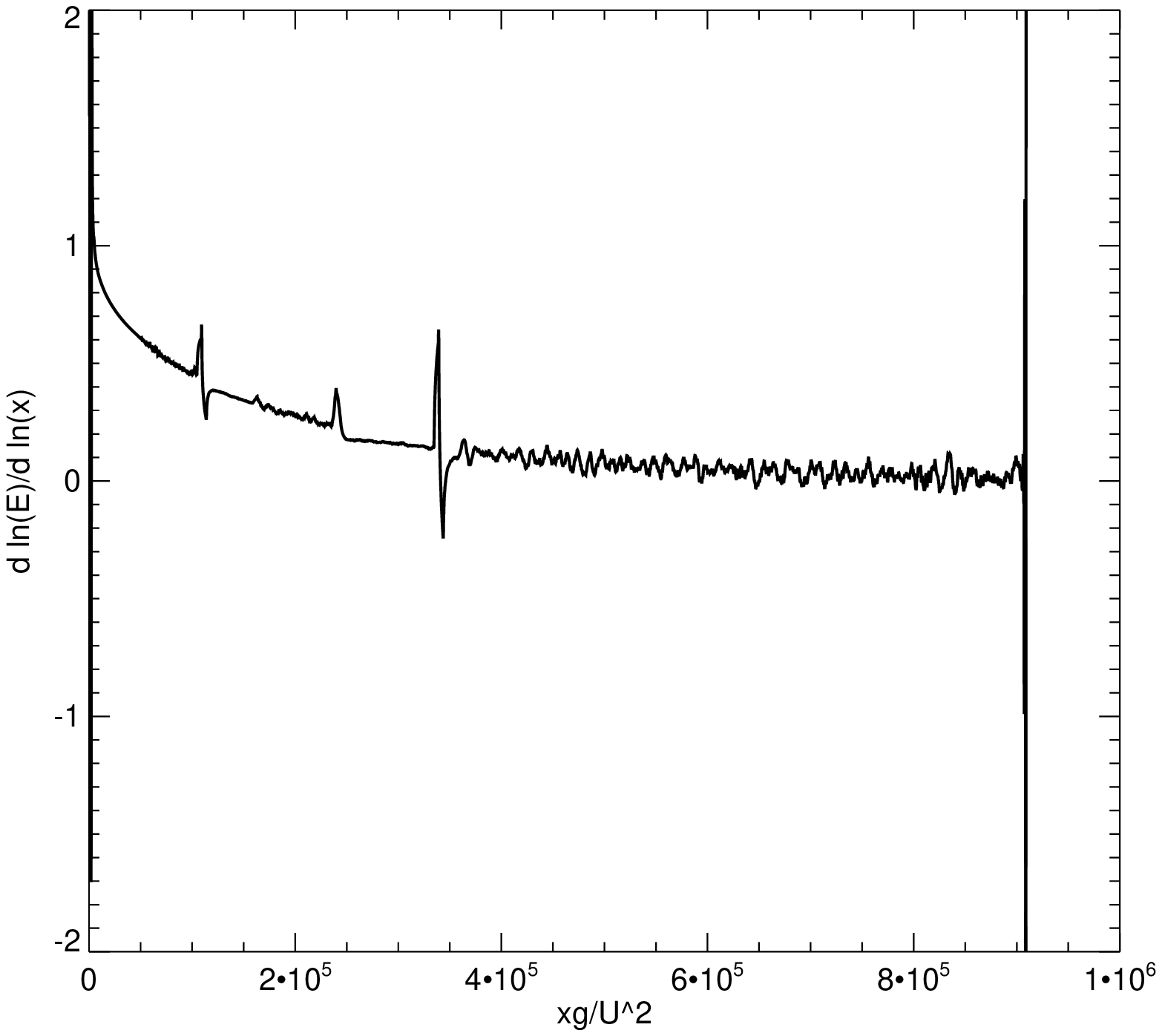}
                \caption{}
                \label{WAM3FormB}
        \end{subfigure}
	\caption{Same as Fig.\ref{AltForm}, but for $WAM3$ $S_{in}$}
	\label{WAM3Form}
\end{figure}

\noindent Fig.\ref{WAM3Form} shows that total energy growth along the fetch dramatically underestimates $ZRP$ simulation, and has the value of exponent $p$ asymptotically going to $0$ versus fetch coordinate $x$.


\begin{figure}
        \centering
        \begin{subfigure}[b]{0.45\textwidth}
                \includegraphics[width=\textwidth]{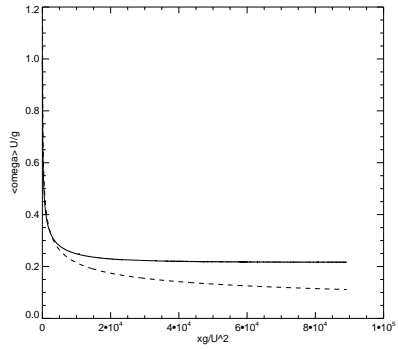}
                \caption{}
                \label{MeanFreqWAM3A}
        \end{subfigure}
\qquad
        \begin{subfigure}[b]{0.45\textwidth}
                \includegraphics[width=\textwidth]{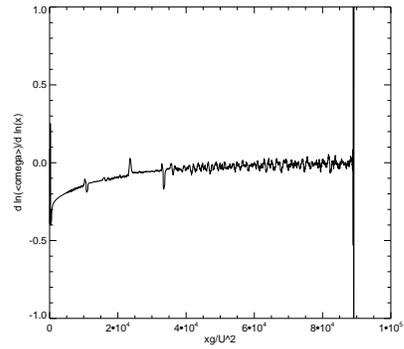}
                \caption{}
                \label{MeanFreqWAM3B}
	\end{subfigure}
	\caption{Same as Fig.\ref{MeanFreq}, but for $WAM3$ $S_{in}$}
	\label{MeanFreqWAM3}
\end{figure}

\noindent Dependence of the mean frequency against the fetch shown on Fig.\ref{MeanFreqWAM3} demonstrates strong discrepancy with $ZRP$ results and corresponding index $q$ also goes to $0$ asymtotically.


\begin{figure}
        \centering
        \begin{subfigure}[b]{0.45\textwidth}
                \includegraphics[width=\textwidth]{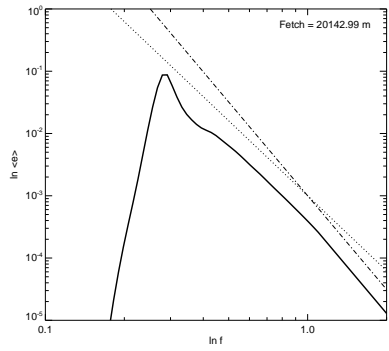}
                \caption{}
                \label{SpecWAM3A}
        \end{subfigure}
\qquad 
        \begin{subfigure}[b]{0.45\textwidth}
                \includegraphics[width=\textwidth]{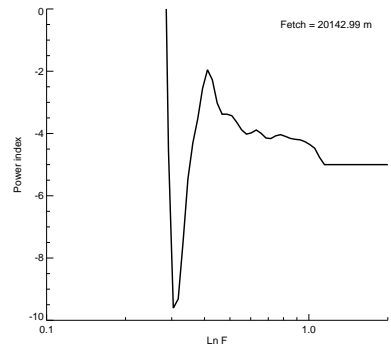}
                \caption{}
                \label{SpecWAM3B}
        \end{subfigure}
        \caption{Same as Fig.\ref{Spec} but for $WAM3$ $S_{in}$}
	\label{SpecWAM3}
\end{figure}

\noindent Fig.\ref{SpecWAM3A} presents directional spectrum as a function of frequency in logarithmic coordinates. One can see:

\begin{enumerate}
\item  the spectral maximum area

\item  Kolmogorov-Zakharov segment $\sim \omega^{-4}$

\item  Phillips high frequency tail $\sim \omega^{-5}$
\end{enumerate}

\noindent Fig.\ref{SpecWAM3B} shows log-log derivative of the energy curve from Fig.\ref{SpecWAM3A}, corresponding to the exponent of the local power law. Again, one can see the areas corresponding  Kolmogorov-Zakharov index $-4$ and Phillips index $-5$. The value of the index to the left side of $-4$ plateau has the tendency to grow, which qualitatively corresponds to the ``inverse cascade'' Kolmogorov-Zakharov index $-11/3$.


\begin{figure}
        \centering
	\includegraphics[width=0.5\textwidth]{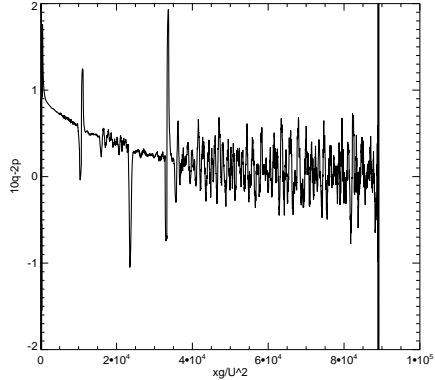}
	\caption{Combination $(10q-2p)$ as a function of the fetch $x$ for $WAM3$ wind input term.}
	\label{MagicNumberWAM3}
\end{figure}

\noindent Fig.\ref{MagicNumberWAM3} presents combination $(10q-2p)$ as the function of the fetch coordinate $x$. It is in total disagreement with the theoretical predictions. There is no any indication of ``magic relation'' Eq.(\ref{MagicLaw}) fulfillment.


\noindent Comparing the results, obtained for Snyder and $WAM3$ wind input terms, we see strong discrepancies. Energy dependence on the fetch for  $WAM3$ model (which is esentially Snyder model with long-wave dissipation) strongly qualitatively differs from Snyder model. When long-waves dissipation is present, the energy is not only much smaller, but also its dependence on the fetch stops to be power-one very soon, and radically differs from all other above considered variants of wind forcing, for which long-wave dissipation is absent.

\section{Conclusion}

\noindent We are offering alternative framework for numerical simulation of $HE$. Being supplied with $ZRP$ wind input term, such approach reproduces the results of more than a dozen of experimental observations.

\noindent We also performed numerical simulations of $HE$ for four another historically well-known wind input terms within the same alternative framework. They demonstrated the results deviating from $ZRP$ simulation.

\noindent To classify the results of the above simulations we applied the set of nonlinear tests to different kinds of wind input terms, and here is the conclusion:

\begin{enumerate}

\item  $ZRP$ forcing term perfectly corresponds to theoretically predicted results like Kolmogorov-Zakharov spectrum $\sim \omega ^{-4} $ , self-similar solutions for energy and frequency with exponents $p=1$ and $q=0.3$ correspondingly,``Magic relation'' $10p-2q = 1$ and reproduces more than a dozen of field experiments. Therefore, it can serve as the benchmark.

\item  All wind input terms pass the test for presence of Kolmogorov-Zakharov law $\varepsilon \sim \omega ^{-4}$. This means that effects of nonlinearity are so strong, that presumably no variation of the wind input term parameterization can suppress it.

\item  $Chalikov$ and $Hsiao-Shemdin$ cases fail $p-$ and $q-$ tests, but pass ``Magic relation'' (quasi-self-similarity) test. 

\item  $Snyder$ case ``approximately'' passes $p-$, $q-$ and ``Magic relation'' tests.

\item  $WAM3$ case fails to pass all exept $KZ-$spectrum test.

\item None of the wind-input parameterization, except $ZRP$ one, can correctly reproduce experimentally observed limited fetch growth.

\end{enumerate}

\noindent In summary, the nonlinearity influence is so robust in the dynamics of $HE$ that one can't ``spoil'' Kolmogorov-Zakharov law $\sim \omega ^{-4}$ for any tested wind input term $S_{in}$. Self-similarity tests like $p-$ and $q-$ tests are the most sophisticated between suggested ones. And the ``magic relation'' test is probably somewhere in-between versus detection of the ``quality'' of particular wind input term. 

\noindent The summary of the tests is presented in Table \ref{Table2}.

\begin{table}
\centering
\begin{tabular}{ | l | l | l | l| l | l|} \hline
\textbf{Experiment} & $p$-test & $q$-test & $KZ$-spectrum & Magic relation & Energy growth \\ \hline 
$ZRP$ & YES & YES & YES & YES & YES \\ \hline 
$Chalikov$ & NO & NO & YES & YES & NO \\ \hline
$Snyder$ & $\approx$ & $\approx$ & YES & YES & NO \\ \hline 
$Hsiao-Shemdin$ & NO & NO & YES & YES & NO \\ \hline
$WAM3$ & NO & NO & YES & NO & NO \\ \hline
\end{tabular}
\caption{}
\label{Table2}
\end{table}

\section{Acknowledgments}

\noindent This research was supported by ONR grant N00014-10-1-0991, NSF grant 1130450 and program of RAS presidium "Nonlinear dynamics in mathematical and physical sciencies". The authors greatfully acknowledge the support of these foundations.

\section*{References}

\bibliography{mybibfile}

\end{document}